 \documentclass[preprint2]{aastex}

\shorttitle{Relaxation of a Collisionless System}
\shortauthors{Merrall \& Henriksen}

\newcommand{\kms}{\hbox{km$\,$s$^{-1}$}}

\begin{document}

\title{Relaxation of a Collisionless System and the Transition to a
New Equilibrium Velocity Distribution}
\author{Thomas E.C. Merrall\altaffilmark{1}}
\affil{Department of Astronomy and Astrophysics, University of Toronto, Toronto, Ontario, M5S 3H8, Canada}

\author{Richard N. Henriksen\altaffilmark{2}}
\affil{Department of Physics, Queen's University, Kingston, Ontario, K7L 3N6, Canada}

\altaffiltext{1}{merrall@astro.utoronto.ca}
\altaffiltext{2}{henriksn@astro.queensu.ca}


\begin{abstract}
In this paper, we present our conclusions from the numerical study of the
collapse of a destabilized collisionless stellar system. 
We use both  direct integration of the Vlasov-Poisson equations and an
N-body tree code to obtain our results, which are mutually confirmed.
We find that 
spherical and moderately nonspherical collapse configurations evolve to new
equilibrium configurations in which the velocity distribution
approaches a Gaussian form, at least in the central regions.  
The evolution to this state has long been
an open question, and in this work we are able to clarify the
process responsible and to support predictions made from statistical
considerations (Lynden-Bell 1967; Nakamura 2000).  The simulations of
merging N-body systems show a transition to
a Gaussian velocity distribution that is increasingly suppressed as
the initial separation of centres is increased.  Possible reasons for this are
discussed.  
\end{abstract}



\section{Introduction}
\label{intro}

Collisionless systems are very important in astronomical contexts and
have been studied extensively in the past (see, for example,
Lynden-Bell 1967; van Albada 1982; Fujiwara 1983; Rasio, Shapiro \&
Teukolsky 1989; Henriksen \& Widrow 1995, 1997, 1999; Kandrup, Mahon
\& Smith 1994; Hozumi, Fujiwara \& Kan-ya 1996; Makino \& Ebisuzaki
1996; Makino 1997; Merritt \& Quinlan 1998; Hozumi, Burkert \&
Fujiwara 2000).  Cold dark matter haloes are considered quintessential
examples of collisionless systems and are treated in the `mean field'
approximation.  In this approximation the individual members of the
system move under the influence of the mean gravitational field of the
entire ensemble, with close-encounters playing only a minor role.

Plasma physicists have also been studying collisionless systems for
most of that field's existence.  One of the most common observations
from plasma physics is the prevalence of single-component Gaussian
velocity distributions.  Since Langmuir's (1925) observations,
Gaussian velocity distributions have been found in laboratory, space,
and simulated plasmas whenever they are supposed to be well-relaxed
(Nakamura 2000); despite having an initial non-Gaussian velocity
distribution.  The Gaussian distribution is so common, in fact, that
very little attention is paid to it.  Experiments and computer
simulations are expected to produce it, and theorists have no doubts
about using it as a starting condition in their calculations (Nakamura
2000). The direct plasma analogy with our problem concerns of course
non-neutral plasma.

We explore in this work the extent to which the same distribution
might be expected to arise in the analogous gravitational problem. We
begin  by looking at common equilibrium systems (polytropes) that we
destabilize by a sudden `cooling'. Their subsequent evolution is
followed by a direct numerical integration of the Vlasov-Poisson
system (referred to herein as the 'CBE integration method') in this
part of the paper and we find eventually a new virialized state. The
central regions of this new equilibrium show a Gaussian velocity
distribution when a top-hat smoothing over the phase-mixing streams is
applied. This work is repeated for several polytropic indices and each
case yields similar conclusions. We detect the phase stream
instability that was found by Henriksen \& Widrow (1997) using a
simple spherical shell code. This two-stream type instability is most
vigorous for `cold' systems and at the current `turn-round' radius
where infalling material interacts strongly with newly turning
material. It appears likely from our calculations that this
instability is one  relaxation mechanism that leads to the
Gaussian distribution. The other, more properly termed `violent
relaxation', involves the bulk time dependence directly as originally
envisaged.    

In the second part of the paper we  use an  N-body tree code to follow
the evolution of a destabilized halo that initially had an equilibrium
velocity distribution. This calculation confirms the CBE integration
in that it shows the same instability leading to the same (Gaussian)
central distribution for sufficiently cold systems (all this in the
absence of a central black hole of course). We also show in this
connection that the oscillations in the virial ratio reported
elsewhere (David \& Theuns, 1989) that appear in these calculations
are finite number effects, as was indeed suggested in Rasio, Shapiro
\& Teukolsky (1989). We show explicitly that these oscillations
decrease in amplitude with increasing N (figure \ref{plumvirial}). The
system behaviour appears to approach the behaviour found in the CBE
integration as $N\rightarrow\infty$. By finite number effects we do
not mean simple statistical fluctuations. Rather it is likely that
these correspond to the wave-particle interaction part of the `violent
relaxation' as discussed in Funato, Makino and Ebisuaki (1992a,b). 

Subsequently in a third part we explore  the relaxation of
merging systems using the tree code. We start the systems (each in
equilibrium) with their centres at various distances and allow them to
evolve under their mutual (all particles active) attraction. The
impact parameter is always zero. We find that there is a distance
beyond which the relaxation to a Gaussian velocity distribution is
suppressed.

In all of the N-body work the finest-grained distribution function $f$ is
found at any time by tagging the particles with the value of $f_o$ at
their initial position. In a collisionless system this is then the
value of $f$ at their current position $(\vec{r},\vec{v})$ and so in
this fashion $f(\vec{r}, \vec{v})$ is found. Coarser grained pictures
may then be obtained by suitable binning of the particles, but no
such smoothing is applied in this part of the paper. These
un-smoothed 
results are consistent with the those found by the CBE integration
method when the latter are smoothed over the phase space streams.    

\subsection{Equilibrium Velocity Distributions: Theoretical
  Background}

The first attempt at the formulation of a statistical analysis of the
Collisionless Boltzmann Equation (CBE) ( also referred to here as
the Vlasov equation since Boltzmann never considered his equation
without collisions) was made by Lynden-Bell (1967).
In this work, Lynden-Bell constructed a statistical theory of equilibrium based on
the conservation of phase-space volume.  The distribution function (DF) obtained by
Lynden-Bell is a superposition of Fermi-Dirac distributions.  In the
nondegenerate limit (defined by Lynden-Bell as the coarse-grained
limit, when the average DF in a macro-cell is much less than the
fine-grained DF), the equilibrium becomes a {\it superposition of
Gaussian components}, with velocity dispersions inversely proportional
to the phase-space density.  This prediction has become known as the
``velocity dispersion problem'' since {\it in the nondegenerate limit
the distribution should simply be a single-component Gaussian, with no
mass segregation in the velocity dispersion}.  Although it suffers
from this problem, Lynden-Bell's paper is seen as a ground-breaking
work in the understanding of the equilibria of collisionless systems.

Following Lynden-Bell, Nakamura (2000) recently attempted to formulate a
statistical theory of a collisionless system which did not exhibit the
velocity dispersion problem.  In his analysis, Nakamura used the
``maximum entropy principle'' of Jaynes (1957a, b).  This principle is
stated as producing, ``the probability distribution over microscopic
states which has maximum entropy subject to whatever is known, [and]
provides the most unbiased representation of our knowledge of the
system'' (Jaynes 1957b).  In this context the term {\it unbiased} is
used in the colloquial sense, without any strict technical meaning.
Jaynes has shown how to construct the standard theory of statistical
mechanics based on this principle, with no need to invoke the ergodic
hypotheses required in standard derivations of statistical mechanics.
Nakamura acknowledges that using this theory simply amounts to making
a different assumption from ergodicity (that of maximum entropy), and
uses it for methodological convenience rather than some innate
conceptual superiority.  

In his paper, Nakamura shows that Lynden-Bell's statistics are
equivalent to calculating the entropy of the system based on the
probability of {\it particle transition} (i.e.~the probability that a
particle at some initial phase-space point moves to some other
point).  This is inconsistent with the entropy calculation of ordinary
collisional gases, in which the entropy is calculated from the
probability of {\it particle existence} (i.e.~the probability that a
particle is located at some point in phase-space).  Nakamura uses the
latter definition in his calculations, and demonstrates how this
difference in entropy calculation can account for Lynden-Bell's
results.  Using the maximum entropy principle, he is able to derive an
expression for the relaxed velocity distribution which is a
single-component Gaussian.  With this result he is able to explain
such phenomena as mass-mixing and the temperature distribution of
solar wind particles (Nakamura 2000).

Additional support to the idea of Gaussian equilibrium velocity
distributions is provided by Henriksen \& Le Delliou (2002),
who developed and studied a new method of coarse graining the
distribution function of a collisionless system.  They note that a
Gaussian distribution is the one that is best-behaved in their
coarse-graining scheme.

\subsection{Gravitational Collapse}
Collisionless systems may start their evolution from a
variety of non-equilibrium initial conditions and they are expected to 
subsequently relax, although the relaxation in energy seems to be more
`moderate' than `violent' (Funato, Makino \& Ebisuzaki 1992a, b). Such
systems have been studied at length using semi-analytic methods and
shell codes  (e.g.~Fillmore \& Goldreich 1984; Bertschinger 1985;
Henriksen \& Widrow 1997; Hoffman \& Shaham 1985; Henriksen \& Widrow
1999) for spherically symmetry and radial orbits.  Some work has also
been done on spherical systems with elliptical orbits (but without net
rotation) as in Sikivie, Tkachev \& Wang (1997) and Henriksen \& Le
Delliou (2002).  However the shell code numerical method generally
lacks the dynamical range and precision in phase space to accurately
follow the evolution and final form of the distribution function.

In this work we perform an integration of the coupled CBE and Poisson
equations based on methods suggested by Cheng \& Knorr (1976),
Fujiwara (1983) and Rasio, Shapiro \& Teukolsky (1989) to remove this
limitation, and so we are able to follow in detail the evolution of
the DF during phase mixing and violent relaxation.

Other work aimed at studying the transition of a disturbed system to its
end state in fully three-dimensional systems has been done using
approximate N-body techniques (van Albada 1982; Funato, Makino \&
Ebisuzaki 1992a, b; Capelato, de Carvalho \& Carlberg 1995; Dantas,
Capelato, de Carvalho \& Ribeiro 2002).  Results obtained by van
Albada (1982), Tanekusa (1987), and Funato, Makino \& Ebisuzaki
(1992a, b) suggest the process of violent relaxation is not sufficient
to take the system to the maximum entropy state predicted by
Lynden-Bell (1967), and Nakamura (2000).

Our present results show that with sufficient initial symmetry it {\it
is} possible for a collisionless system to relax to a Gaussian
velocity distribution.  Other aspects of our results are in agreement
with observations made by the above authors -- for example, we agree
with the correlation in particle energies between initial and final
states of collapsing dissipationless systems (moderate 'violent
relaxation').

\section{Mathematical Formulation}
\label{math_form}
We have used two different methods of calculation in this paper.  For the
investigation of a spherically symmetric collapse, rather than
using approximate N-body methods (e.g.~Barnes \& Hut 1986 and methods
derived from it), we have integrated the CBE directly. Typical
approximate N-body simulation techniques do amount to solving the CBE
along a finite set of particle trajectories, with subsequent smoothing
over many particles to find physical quantities (see e.g.~Quinn
2001). The CBE integration method differs however in that it permits
us to examine essentially any set (however large) of particles we
wish, with the DF automatically conserved along each particle
trajectory.  This eliminates the need to smooth over many particles,
as the value of $f$ is found at any phase-space point simply by the
integration process.

For the nonspherical collapse simulations, however, it was necessary
to employ approximate N-body techniques to calculate the DF evolution,
as described below (see section \ref{nbodcomp}).

\subsection{The Collisionless Boltzmann Equation}
The CBE is a statistical equation which uses a distribution function
to describe how an ensemble of self-gravitating but otherwise
non-interacting particles will behave.  The probability distribution
function is defined as, 
\begin{equation}
\label{DF_def}
f\!\left({\bf x}, {\bf v}\right) = \frac{d^6N}{d^3{\bf x}\, d^3{\bf
v}}\; . 
\end{equation}
This represents the number of particles inside some differential
phase-space volume.

The CBE is given by, 
\begin{equation}
\label{full_CBE}
\frac{\partial f\!\left( {\bf r}, {\bf v}, t \right)}{\partial t} + {\bf
v} \frac{\partial f\!\left( {\bf r}, {\bf v}, t \right)}{\partial {\bf
r}} + {\bf a}\!\left[ f \right] \frac{\partial f\!\left( {\bf r}, {\bf
v}, t \right)}{\partial {\bf v}} = 0 \; ,
\end{equation}
with ${\bf v}$ and ${\bf a}\left[ f \right]$ representing the velocity
and acceleration respectively.  The acceleration is a functional of
$f$ and can be calculated directly from it at any point in space, and
at any time.  The velocity is an independent phase-space direction;
together with the position an orthogonal basis is defined. 

Without making any further restrictions the CBE can, in principle, be
integrated in the full six-dimensional (e.g.~$x, y, z, v_x, v_y, v_z$)
phase space.  This is, unfortunately, prohibitively time consuming.
For this reason, we must impose certain restrictions.  We must either
treat the system as a discrete system of particles and make
approximations to the gravitational force (the most common
approach), which leads to statistical fluctuations and necessitates
smoothing over many particles to obtain interesting information
(e.g.~density), or restrict our system through the use of symmetries.
This section uses the latter method to investigate the gravitational
collapse of a stellar polytrope. 

For the current application to a system that is constrained to
maintain spherical symmetry, equation (\ref{full_CBE}) is reduced to a
dependence only on $r$, $v_r$, and $j^2$ following Fujiwara (1983).
It now reads,
\begin{equation}
\label{CBE}
\frac{\partial f}{\partial t} + v_r\frac{\partial f}{\partial r} + \left(
\frac{j^2}{r^3} - \frac{G \, M\!\left(r\right)}{r^2} \right) \frac{\partial
f}{\partial v_r} = 0 \; .
\end{equation}
This represents the evolution equation of a system which is
constrained to remain spherically symmetric with no net rotation, but
with constituent particles that can have nonzero angular momentum
(i.e.~the particles are not limited to radial orbits, and the angular
momentum vectors of the particles are uniformly distributed).  

\subsection{Calculation using the distribution function}
With knowledge of the spherically-symmetric DF we are able to
calculate physical quantities by integration of the DF. The
density, kinetic energy, and potential energy are given by, 
\begin{equation}
\label{rho}
\rho\left( r \right) = \frac{\pi}{r^2} \int_{-\infty}^{\infty}\!\!\!
dv_r \int_{0}^{\infty}\!\!\! dj^2 f\!\!\left( r, v_r, j^2 \right) \; , 
\end{equation}
\begin{equation}
\label{KE}
T = 2 \pi^2 \int_{0}^{\infty}\!\!\! dr \int_{0}^{\infty}\!\!\! dj^2
\int_{-\infty}^{\infty}\!\!\! dv_r \left( v_r^2 + \frac{j^2}{r^2} \right)
f\!\!\left( r, v_r, j^2 \right) \; , 
\end{equation}
\begin{equation}
\label{PE}
W = 2 \pi \int_0^{\infty}\!\!\! \rho\!\left( r \right) \Phi\!\left( r
\right) r^2 \, dr \; . 
\end{equation}
The mass distribution $M\!\left( r \right)$ for a known or calculable
DF can be calculated by finite-differencing the spherically-symmetric
potential,
\begin{equation}
\label{Mfinite}
M\!\left( r_i \right) = i^2 \, \delta r \left(\frac{\Phi_{i - 1} - \Phi_{i +
1}}{2 \, G}\right) \; .
\end{equation}
In this equation, $i$ labels the radial grid point, $\delta r$ is the
spacing between points on the potential grid, and $\Phi_i$ is the
calculated potential on the grid.

Using the above equations ((\ref{rho}) -- (\ref{Mfinite})) we are able
to calculate the density, potential, and mass distributions directly
from the DF, as well as monitoring the total energy ($T\!+\!W$) and
virial ratio ($-2 T/W$) at each timestep.

\subsection{Polytropic distributions}
We have chosen to assume an initially polytropic distribution for part
of this investigation.  A polytrope has a DF which is simply a
power-law in energy, $f\!\left( E \right) \!\propto\! \left( -E
\right)^{n - \frac{3}{2}}$ (Camm 1952; Binney \& Tremaine 1987), with
a gravitational potential that is flat (i.e.~no cusp) in the center.
With these two conditions and the polytropic index, $n$, we are able
to uniquely construct the phase-space matter distribution.  Models
with power-law dependence on energy and cusped central densities were
previously studied by Henriksen \& Widrow (1995).

The initial DF is then taken as, 
\begin{equation}
\label{DF}
f = \left\{ \begin{array}
{r @ {\quad:\quad}l}
F \left( -E \right)^{n - \frac{3}{2}} & E < 0 \\ 0 & E \ge 0
\end{array} \right. \; ,
\end{equation}
where $n$ is the polytropic index, and the energy is given by, 
\begin{equation}
\label{energy}
E = \frac{1}{2} \: \alpha^2 \! \left( {v_r}^2 + \frac{j^2}{r^2} \right) +
\Phi \left( r \right) \; . 
\end{equation}
The coefficient of the DF, $F$,  is chosen so the total mass is
normalized to unity, and for general $n$ is given by,
\begin{equation}
\label{poly_coeff}
F = \frac{\Gamma\!\left(n+1\right) \, A^2 \, \alpha^{3}}{4 \pi \left(2
\pi \right)^{\frac{3}{2}} \Gamma\!\left( n - \frac{1}{2} \right)} \; , 
\end{equation}
with $A^2 = 4 \pi G \rho_c$ depending on the order of the polytrope.
This quantity is commonly tabulated for several values of $n$
(q.v.~Chandrasekhar 1939).

The parameter $\alpha$ is a measure of the initial stability of the
system.  In the calculation of the DF, velocities are reduced with the
substitution $v_r \!\to\! \alpha v_r$, $j \!\to\! \alpha j$.  As
$\alpha$ is increased, the value of $f$ is calculated assuming larger
values of $v_r$ and $j^2$ (and therefore, larger kinetic energy) than
are actually present.  This means that for a given value of the DF the
velocities are decreased from what they would be in a stable
configuration by a factor $1/\alpha$, so the sphere loses thermal
support and is free to collapse under the influence of its
self-gravity until it reaches a new equilibrium.  Since the velocities
are all decreased by this constant factor, the kinetic energy must be
similarly decreased ($K\!\to\!K/\alpha^2$) with the potential energy
remaining unchanged, and the  initial virial ratio of the system is
reduced to $1/{\alpha}^2$.  Of course, if $\alpha$ is set equal to
unity, the distribution is in virial equilibrium and will not
collapse.  This has been numerically verified (see section
\ref{poly_results_section}), which demonstrates that the initial
polytrope is indeed stable to small perturbations. 

In order to begin our calculation, we must find the initial potential
of the distribution by integrating the Lane-Emden equation (see
e.g.~Binney \& Tremaine 1987; Kippenhahn \& Wiegert 1990;
Chandrasekhar 1939), 
\begin{equation}
\label{LE_scaled}
\frac{d^2 w}{d z^2} + \frac{2}{z} \, \frac{d w}{d z} = -w^n \; , 
\end{equation}
with the boundary conditions $w(0) = 1$, $w'(0) = 0$.  For general
$n$, no analytic solution exists so the potential must be found by
solving (\ref{LE_scaled}) numerically.

Once the potential is found, we are able to construct the initial DF
from equations (\ref{DF}) -- (\ref{poly_coeff}).  This allows us
subsequently to self-consistently evolve the distribution under the
influence of its self-gravity by solving the CBE as described below.

\subsection{N-Body simulation}
\label{nbodcomp}
While the CBE integration method detailed above is able to provide
accuracy, it can also be computationally intensive in systems of lower
symmetry.  Consequently, we have used a
treecode (Barnes \& Hut 1986) (modified to carry the initial DF along
with each particle) to simulate the collapse of a system without
requiring it to remain spherically symmetric.  The initial conditions
used in the simulations were provided by the Kuijken \& Dubinski
(1995) three-component galaxy software detailed below.

\subsection{Lowered Evans Haloes}
\label{evanshalo}
For this investigation, we used a halo given by a lowered Evans model
(Kuijken \& Dubinski 1994, 1995 -- hereinafter KD94, KD95).   This
model provides a halo with a finite mass, adjustable core radius, and
a flat rotation curve over an appreciable amount of its extent.  The DF
for this model is given in KD94, and is reproduced here, 
\begin{equation}
\label{KD_DF}
f \!\left(E, L_z\right) \!=\! \left\{ \!\!\!\begin{array}
{r @ { \quad}l}
\left[\left(A_{\mbox{o}}
L_z^2+B_{\mbox{o}}\right) \mbox{exp}\!\left(-E/\sigma_0^2\right) +
C_{\mbox{o}}\right]&\\  
\times \left[\mbox{exp}\!\left(-E/\sigma_0^2\right)-1\right]&
\mbox{if } E\! < \!0\\ 
& \\
0 & \mbox{if } E\! \ge \!0 
\end{array} \right.
\end{equation}
The coefficients $A_{\mbox{o}}$, $B_{\mbox{o}}$, and $C_{\mbox{o}}$
are given by, 
\begin{eqnarray}
\label{KDABC}
A_{o} &=& \frac{\left(1 - q^2\right) G
\rho_1^2}{\sqrt{\pi}q^2\sigma_0^7}\nonumber\\ 
B_{o} &=& \frac{4 R_c G \rho_1^2}{\sqrt{\pi}q^2\sigma_0^5}\\
C_{o} &=&
\frac{\left(2q^2-1\right)\rho_1}{\left(2\pi\right)^{3/2}q^2\sigma_0^3}\nonumber
\; .
\end{eqnarray}
Spherically symmetric models have $A_{\mbox{o}}=0$.  Coreless models
have $B_{\mbox{o}}=0$.  When $A_{\mbox{o}}=B_{\mbox{o}}=0$, we
recover King's model (King 1966).

The Evans halo was generated using software written by Kuijken \&
Dubinski (KD94, KD95).  The halo portion of the KD94 \& KD95 software
takes five parameters (one of which is constrained to unity in a pure
halo model).  These parameters are, $\Psi_0$ -- the central potential;
$v_0 = \sqrt{2}\sigma_0$ -- the central velocity scale ($\sigma_0$ is
the central velocity dispersion); $q$ -- an optional flattening
parameter; $(r_c/r_k)^2$ -- the ratio of the core radius to the King
radius (this is constrained to unity for halo-only models such as the
one we are using); and $R_a$ -- a halo scaling parameter (this is
roughly the radius at which the halo rotation curve would reach the
value $\sqrt{2}\sigma_0$ if it were continued at its $R=0$ slope).
These are all scale-free parameters (with Newton's $G$ assumed equal
to unity) and can be rescaled according to our needs.  The physical
scalings used in this work are presented below.

\section{Computational Method}
\label{rst_int}
\subsection{CBE integration}
The evolution of each polytropic system was followed using the CBE
coupled with Poisson's equation to provide the self-gravity.  There
have been several works (e.g.~Watanabe et al.\ 1981; Nishida
et al.\ 1981; Fujiwara 1983; Hozumi, Burkert \& Fujiwara 2000) which
make use of Cheng \& Knorr's (1976) `operator splitting' method of
integrating the CBE in which the DF is alternately evolved through a
`free-streaming phase' during which the acceleration is taken to be
zero for one half-timestep, and an `acceleration phase' during which
forces on the pseudoparticle are accounted for.  A pseudoparticle is
placed at the point of interest, (${\bf r_n}$, ${\bf v_n}$, $t_n$),
and integrated back to the {\it previous} time position, (${\bf
r_{n-1}}$, ${\bf v_{n-1}}$, $t_{n-1}$).  The pseudoparticle will, in
all likelihood, not land exactly on a grid point and so $f$ must be
interpolated at that point from the values at the nearby grid
points. By Liouville's theorem, $f$ is conserved along the trajectory,
$\frac{Df}{Dt} = 0$.  This method is accurate to second-order in time,
although it has the drawback of requiring an Eulerian grid in the
phase-space which can cause amplification of error by the repeated
interpolation of the DF back to the grid at each timestep.  In
addition, when the DF experiences a large degree of phase mixing,
details on scales smaller than the grid spacing will be washed out
(see Rasio, Shapiro \& Teukolsky (1989) for more details). 

For this investigation we have implemented a scheme proposed in Rasio,
Shapiro \& Teukolsky (1989), in which the conservation of $f$ is
again used explicitly.  Since the mass distribution is known for all
times previous to that being calculated, the acceleration for all
times can be calculated.  It is then possible to follow a trajectory
backwards in time to $t \!=\! 0$.  By tracing a pseudoparticle from
the point of interest, (${\bf r}_n$, ${\bf v}_n$, $t_n$), to its point
at the initial time, (${\bf r}_o$, ${\bf v}_o$, $t_o$), the new DF can
be calculated.  The result of this integration tells us where in
phase-space the particle must have {\it started from} at $t \!=\! 0$
to have arrived at the specified point at $t \!=\! t_n$.  Once again,
the DF is conserved along this trajectory, so $f\left({\bf r}_n, {\bf
  v}_n, t_n\right) = f\left({\bf r}_o, {\bf v}_o, t_o\right)$.  This
allows us to use the value of $f$ initially specified, and avoid the
problem of compounding the repeated interpolation error.  The error in
$f$ at any time is essentially independent of that at any other time,
and we are free of the problem of compounding errors.

The integrals for equations (\ref{rho})--(\ref{PE}) were calculated
using a multidimensional adaptive quadrature.  This allows a required
accuracy to be specified, with the software adding interior points to
the integration domain until this accuracy is reached.  This adaptive
method is necessary to handle the phase-separation that appears during
a cold (large $\alpha^2$) collapse, when the streams do not fill a
large fraction of the available phase-space.

The polytrope collapse simulations were performed in scaled variables
with $R = r/a$, $V_r = v_r / \sqrt{G M_{tot}/a}$, $J = j / \sqrt{G
M_{tot} a}$, and $T = t / \sqrt{a^3 / (G M_{tot})}$.  In these cases,
the physical quantities are scaled by what will be referred to as
`characteristic' quantities (e.g.~the characteristic time is $t_o =
\sqrt{a^3 / (G M_{tot})}$).  It is then easy to extract physical
quantities by choosing a total physical mass, $M_{tot}$, and a length
scale, $a$.  With these dimensional scalings, the half-mass crossing
time for an $n=5$ polytrope is $T_{1\!/\!2} = 2R_{1\!/\!2}/\sigma_o
\sim 1.7 t_o$.  

\subsection{N-body integration}
The gravitational force for the nonspherical collapse was calculated
using a treecode (Barnes \& Hut 1986).  This method saves a great
deal of computational effort over the na\"{\i}ve direct pairwise
summation approach by using an approximation to calculate the force
due to distant particles.

The main item of interest in this paper is the expected transition of the
velocity distribution to a Gaussian form during the collapse.  As with our
CBE integration scheme, we have made use of the conservation of the DF
directly by tagging each N-body particle with its initial DF,
calculated from equation (\ref{KD_DF}).  We can then simply plot that
value against the particle velocity to gain an accurate visualization
of the transition of the fine-grained DF to its final form. As with
Henriksen \& Le Delliou (2002), we would normally interpret a ``fully
relaxed'' state to exist when the fine-grained and coarse-grained DFs
coincide. However at the `finest' (individual particle) scale the
statistical treatment itself breaks down so that we can expect to have
to apply some reasonable smoothing to the N-body calculation in order
to define even a fine-grained DF. In this sense coarse-graining is
merely a matter of degree. We give our results for the DF however
simply by plotting each particle in velocity space. Thus no smoothing
is applied except by way of the Gaussian fit. 

\section{Results}
\subsection{Spherically symmetric polytropes}
\label{poly_results_section}
\subsubsection{Test-bed simulations}
The spherically symmetric CBE integration software was tested on
several stable polytropic distributions.  All stable polytrope
simulations were allowed to run for nine characteristic times.
Throughout the lifetime of the simulations, the virial ratio, total
energy, mass profile, and density profile were accurately conserved.

\subsubsection{Collapse simulations}
Once the test-bed simulations above had convinced us of the accuracy
of our code, we performed several collapse simulations.  The spheres
were destabilized by increasing $\alpha^2$ from unity. This
effectively reduces the kinetic energy  by a factor of $\alpha^2$,
and so also the initial virial ratio.

\begin{figure}
\plotone{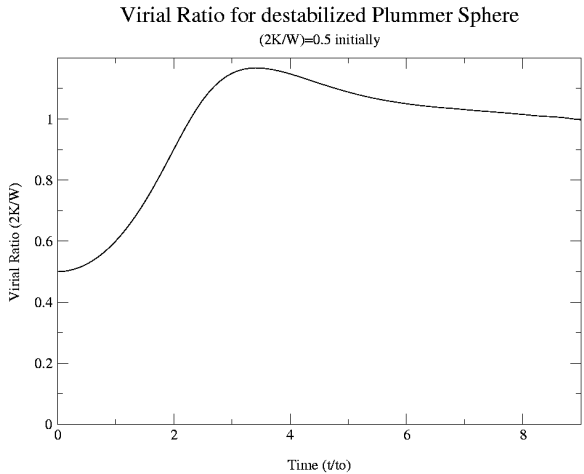}
\caption{\label{VR_n5_a2} 
Evolution with time of virial ratio for $n = 5$ polytrope with initial
virial ratio 0.5.  The virial ratio peaks and approaches unity, with
the system essentially at virial equilibrium within eight
characteristic times.}  
\end{figure}

\begin{figure}
\plotone{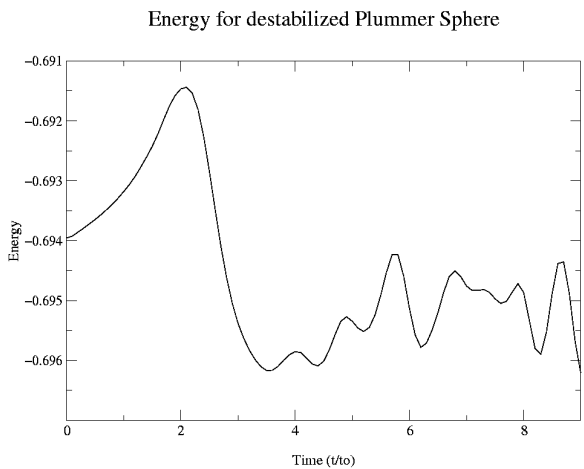}
\caption{\label{E_n5_a2} 
Evolution with time of energy for $n = 5$ polytrope with initial
virial ratio 0.5.  Over nine characteristic times, the energy is
conserved to within 0.5\%.} 
\end{figure}

Results for the $n = 5$ case are shown in Fig.\ (\ref{VR_n5_a2}) --
(\ref{E_n5_a2}).  
Starting from $\left|2K/W\right| = 0.5$, we can see
that the virial ratio peaks above unity, then decreases toward that
value. We note the absence of vigorous oscillations in the virial
ratio about the equilibrium value.  

At first glance the absence of oscillations may appear to
contradict the results of David \& Theuns (1989), who observed
long-lived radial pulsations in $N$-body collapse simulations of
homogeneous spheres.  However the results shown in Fig.\
(\ref{VR_n5_a2}) confirm those seen by Rasio, Shapiro \& Teukolsky
(1989) (their fig.\ 2). Those authors conclude that the adaptive
integration method produces results which approximate an $N$-body
integration as $N\to\infty$, and so suppresses the virial oscillations
which likely result from finite-number effects.
 
We have tested this supression using the N-body treecode to calculate
the virial ratio for the same case as in Fig. \ (\ref{E_n5_a2}) using 
various particle numbers with the tree code that is used below. The
direct integration calculation is superimposed. We see clearly that
the increasing number of particles reduces the amplitude of the
oscillation. Only very weak and smooth oscillations remain in the
largest number used. The direct integration accurately reproduces this
large $N$ behaviour as far as the calculation could be continued. It
is likely that subsequent behaviour of the direct integration would
show even weaker oscillation, but we are unable to demonstrate this
because of the time requirements of the CBE code. 

\begin{figure}
\plotone{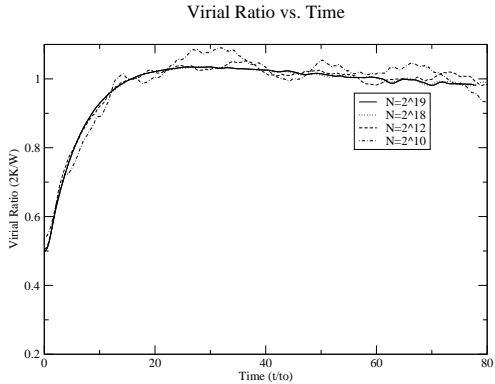}
\caption{\label{vr_many} 
Evolution with time of virial ratio for a self-gravitating system of
particles with initial virial ratio 0.5.  Oscillations and numerical
artifacts are suppressed as the number of particles is increased.
This supports the assertion that the direct integration method does,
in fact, allow us to approach the continuum limit.} 
\end{figure}

A subsequent test was performed using the N-body treecode by
repeatedly evolving a destabilized galactic halo (as decribed in
section \ref{evanshalo}) with initial virial ratio of $0.5$ with 
different numbers of particles. We are able to run this code over a
much larger number of relaxation times than is the case for the CBE
integration. These results are reported in Fig.\ (\ref{vr_many}) where
we see in fact that the virial oscillation amplitude decreases as the
particle number is increased. 
We conclude that not only are {\it any
virial oscillations appearing in this type of simulation likely due
to finite number effects} but that in addition  the results of this
section are consistent with those obtained through traditional
$N$-body methods in a significantly smaller number of relaxation times. 

\begin{figure}
\plotone{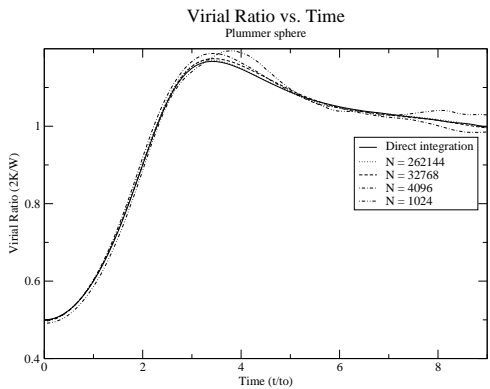}
\caption{\label{plumvirial}
In this figure we show the virial ratio for the n=5 polytrope (Plummer
Sphere) calculated using the tree code for a variety of particle
numbers. The direct integration calculation is superimposed. The
finite number effects are clearly visible, and only a very weak smooth
oscillation remains with the largest number of particles.}
\end{figure}

Having established empirically that the virial oscillations are most
visible with a smaller number of particles in the system, and that the
direct integration approaches the infinite particle limit, we must 
nevertheless be careful not to dismiss the oscillations as merely
errors. There could be finite number effects that are physical and
indeed those observed do exceed the $1/\sqrt{N}$ fluctuation noise (we
are indebted to a referee for this remark). It is possible that we are
seeing the wave-particle aspect of violent relaxation (Funato, Makino
and Ebisuzaki 1992a,b) best in the small systems. For in these systems
the short wavelength time disturbances will be suppressed due to lack
of numerical resolution, leaving only the more global oscillations and
relaxation. And in support of this idea we see that the oscillations
in Fig. \  (\ref{plumvirial}) have periods of several crossing times
in the most pronounced case.   
As the number of particles increases,
the short wavelengths should be progressively filled in and the
relaxation will become more complete on all scales. This is probably
the import of the observed stabilization of the virial ratio with
increasing $N$ and in the direct integration limit. 

Thus rather than being considered as errors the oscillations in small
$N$ systems should be seen as providing a means of studying the actual
development of the collisionless relaxation. It is likely that the
onset of the phase mixing instability that we report is only evident
with sufficiently large $N$ for example. However this is not the main
interest of the present work, wherein we choose rather to study the
evolution of the DF.
  
Returning to Fig.\ (\ref{VR_n5_a2}) which should represent the large
$N$ limit as argued above, we conclude that after approximately eight
characteristic times the cloud is almost fully virialized ($2K/W =
1.01$).  Using values for a dark matter halo obtained through
observations of satellites of the Milky Way (M$_{halo} \sim
1.3\times10^{12} $M$_\odot$, R$_{halo} \sim \;160$ kpc) (Little \&
Tremaine 1987; Arnold 1992) to scale our results, one finds that the
system has almost fully relaxed in approximately $6.6$ Gyr.   

\begin{figure}
\plotone{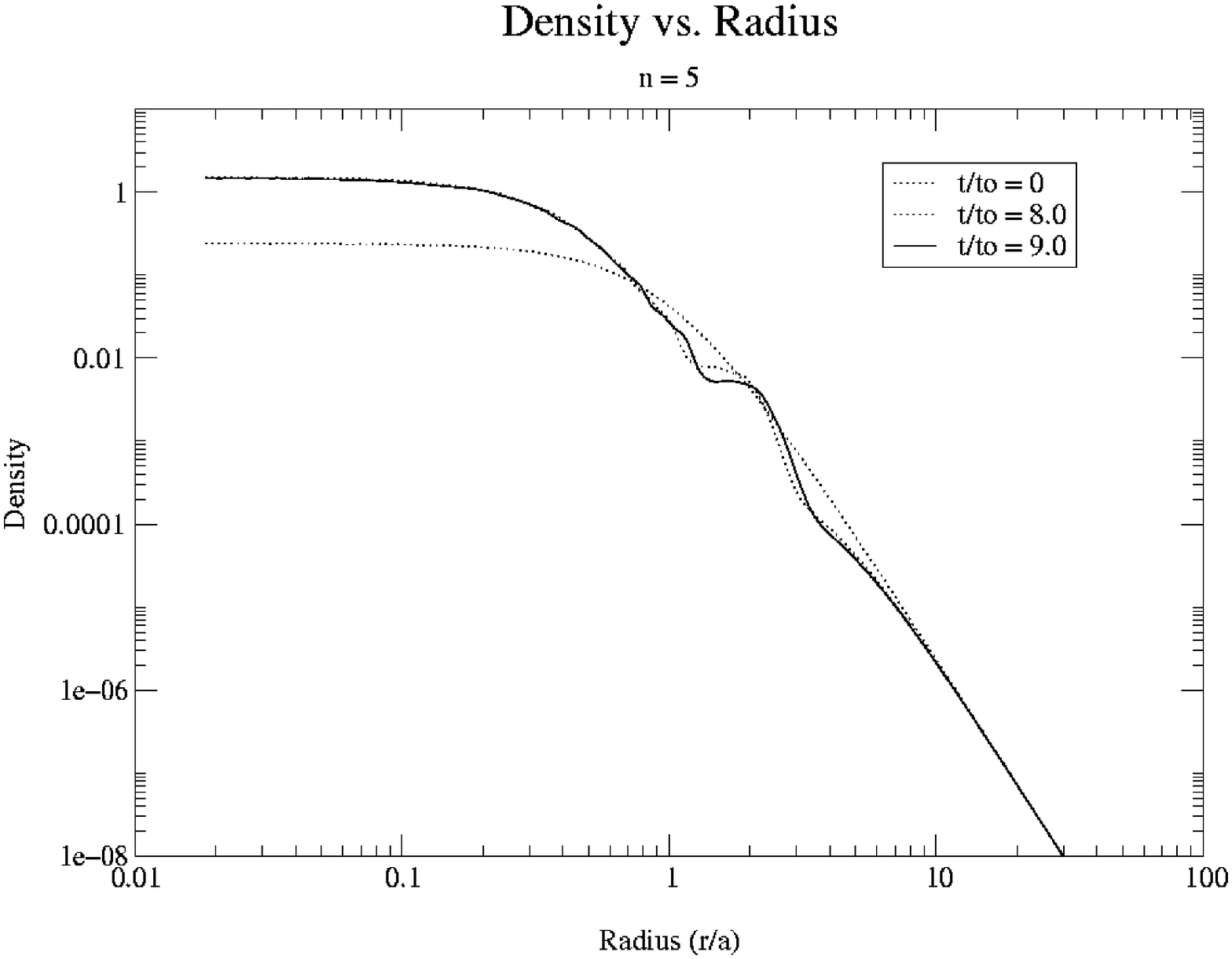}
\caption{\label{rho_n5_a2} 
Evolution with time of density for $n = 5$ polytrope with initial
virial ratio 0.5.  Density waves are clearly seen propagating outward.} 
\end{figure}

\begin{figure*}
\begin{tabular}{|c|c|}
\rotatebox{270}{\scalebox{0.30}
        {\includegraphics{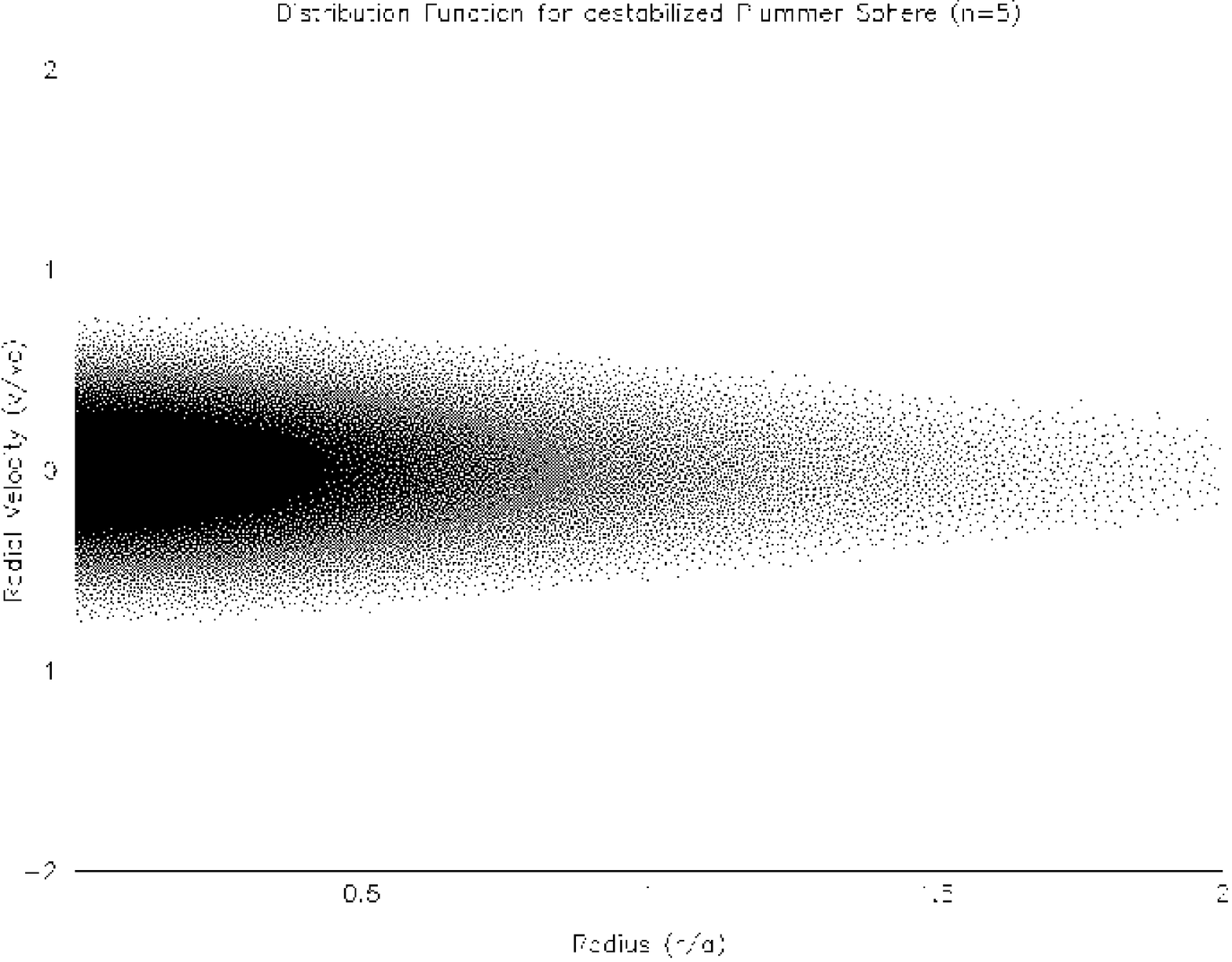}}} & 
\rotatebox{270}{\scalebox{0.30}
        {\includegraphics{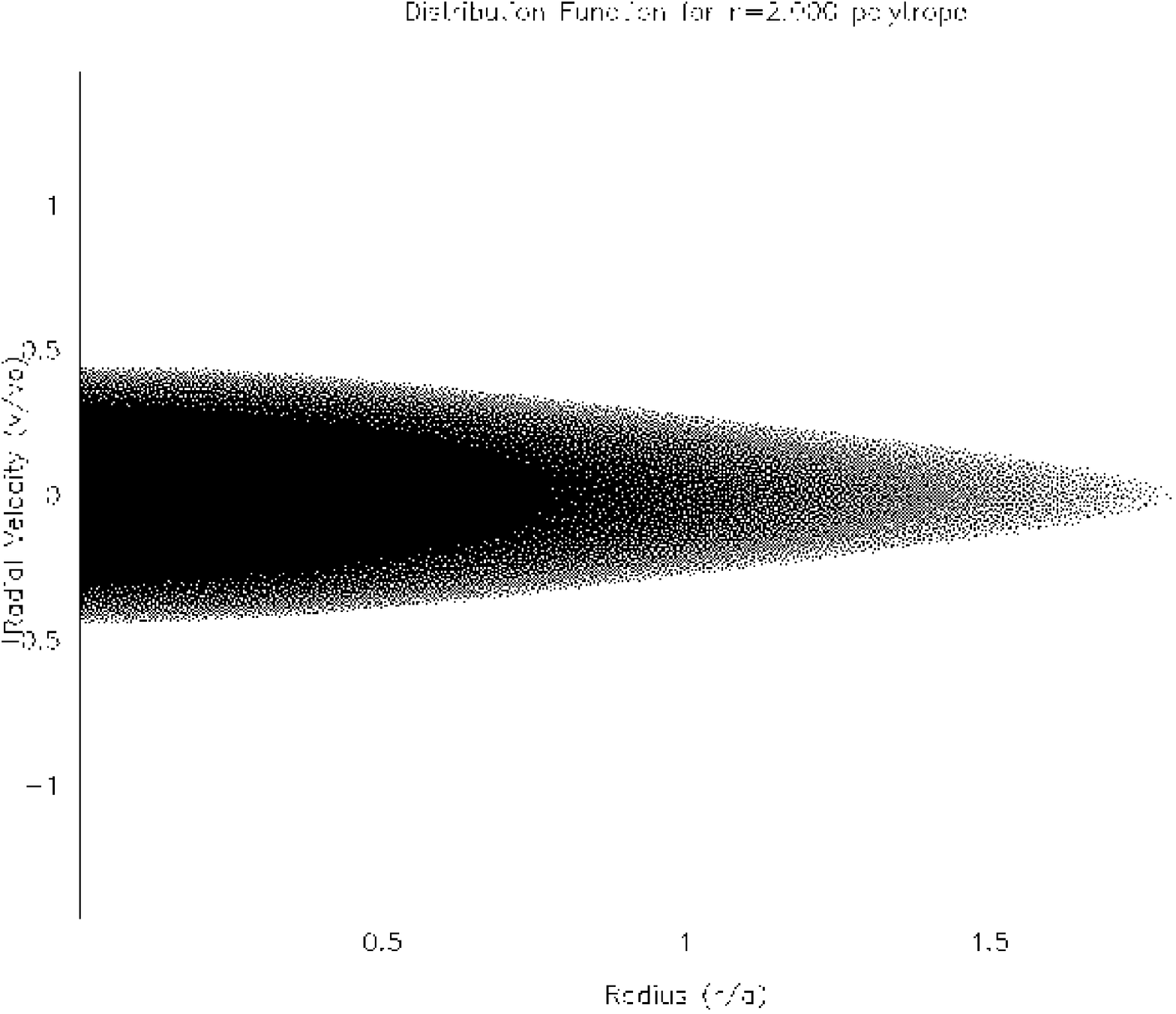}}} \\
\rotatebox{270}{\scalebox{0.30}
        {\includegraphics{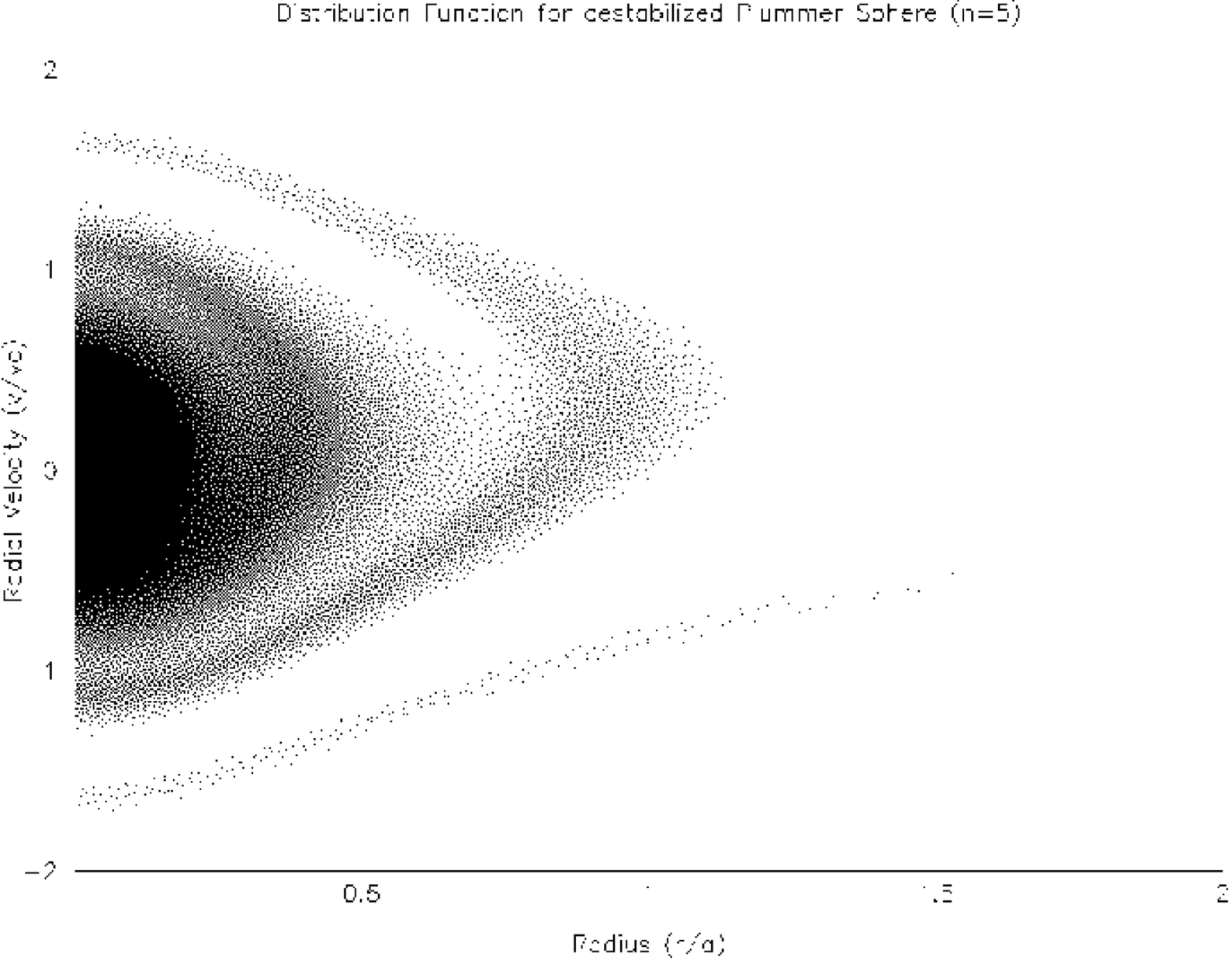}}} & 
\rotatebox{270}{\scalebox{0.30}
        {\includegraphics{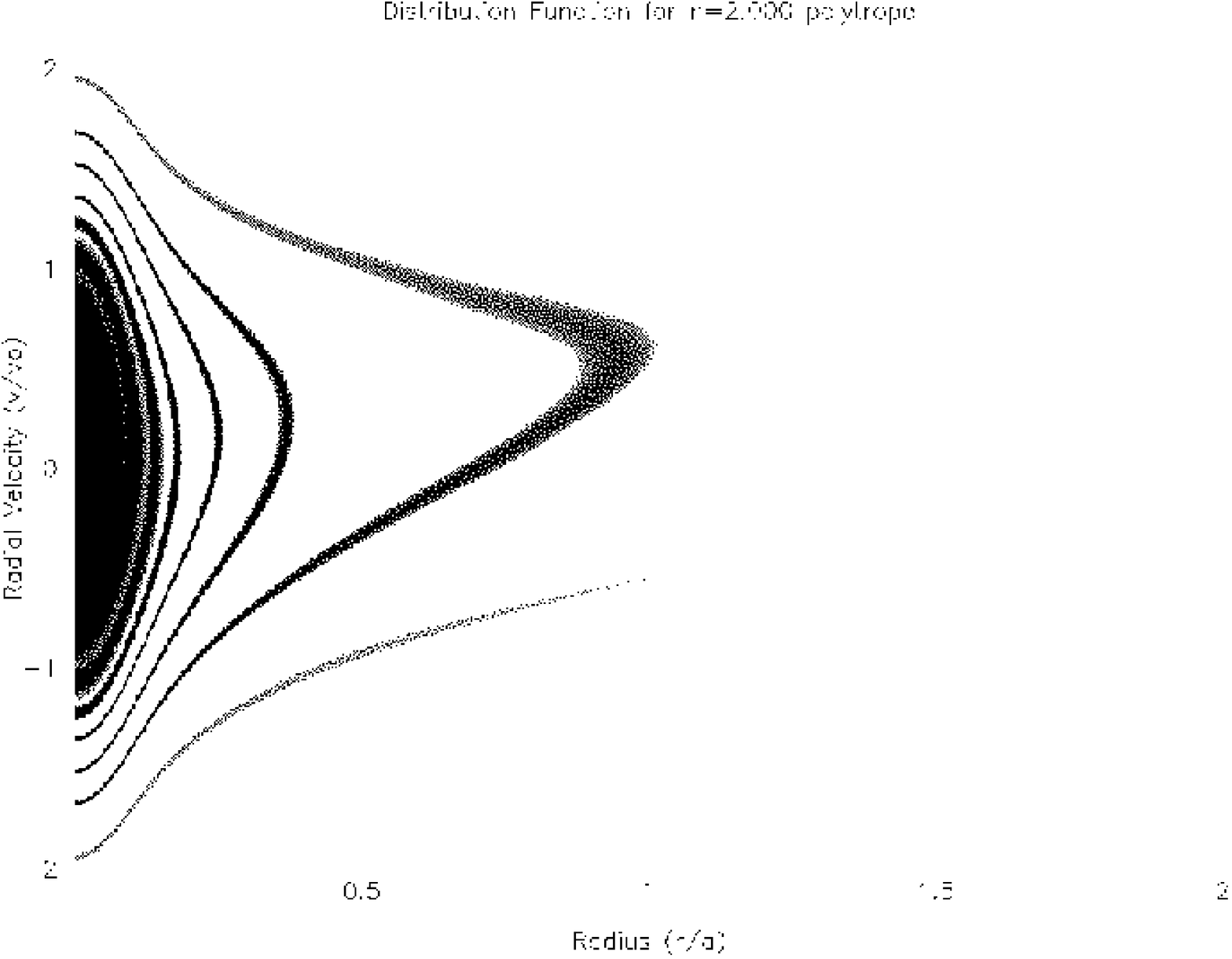}}} \\
\rotatebox{270}{\scalebox{0.30}
        {\includegraphics{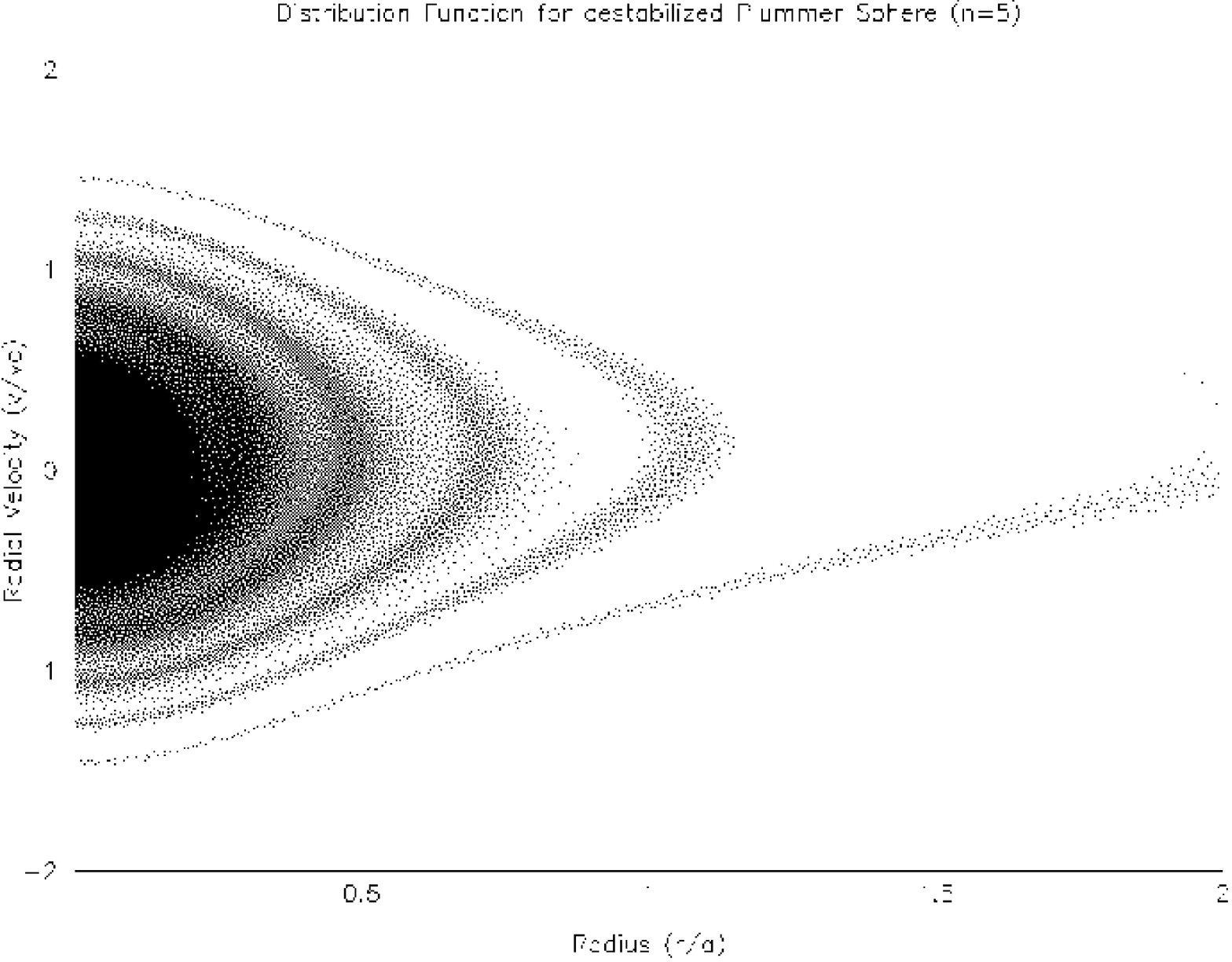}}} & 
\rotatebox{270}{\scalebox{0.30}
        {\includegraphics{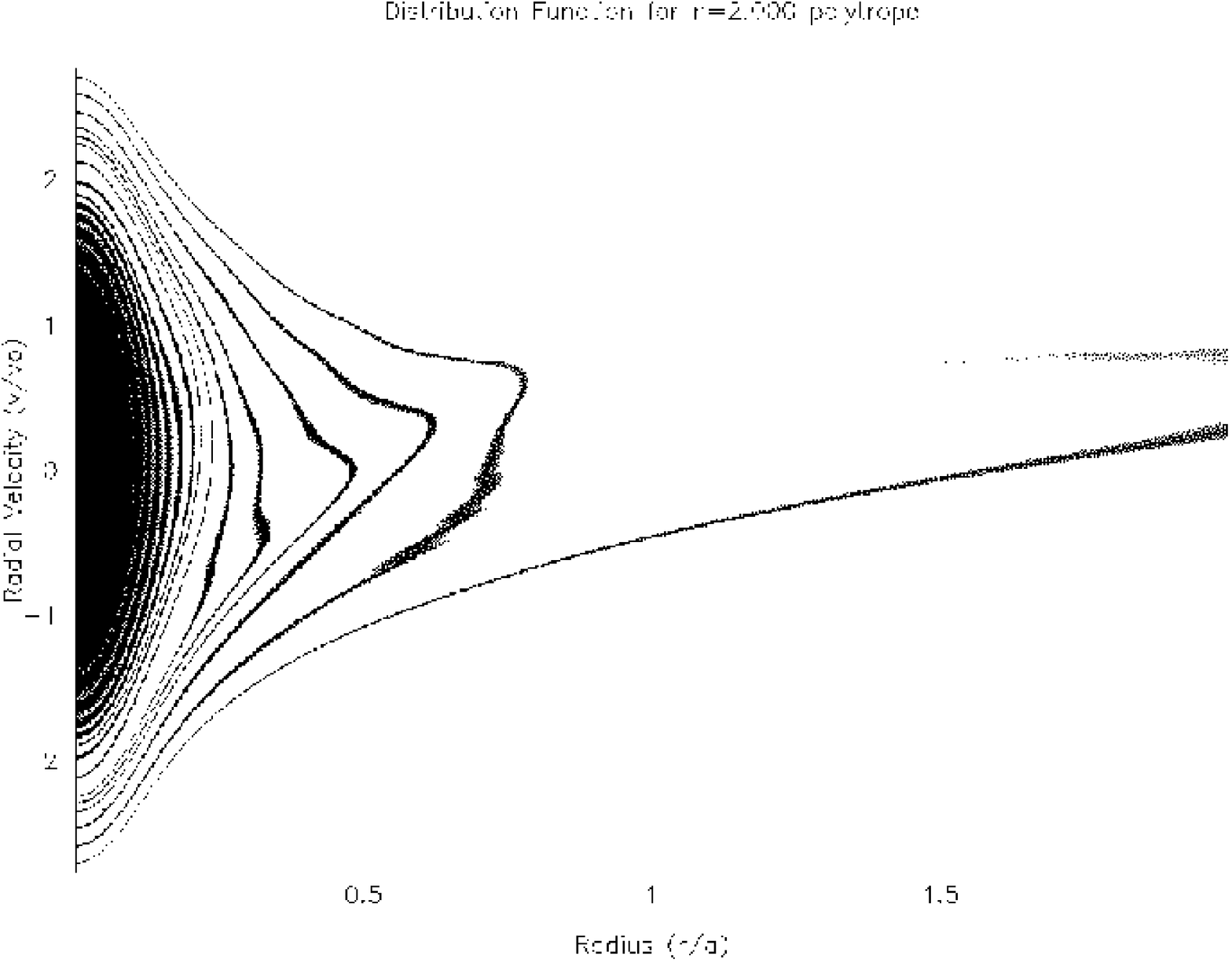}}}
\end{tabular}
\caption[Phase portraits for $n = 5$ and $n = 2$]{\label{pdf_all} 
$j=0$ slices of phase space portrait for $n = 5$ (left column) and $n =
2$ (right column) polytrope with initial virial ratio 0.5 and 0.1
respectively.  Times shown are (from top) $t/t_o = 0.0,\, 4.0,\, 8.0$.}
\end{figure*}

The density profile of the virialized cloud (Fig.\ (\ref{rho_n5_a2})),
shows that the central core has contracted and the density has
increased with respect to its initial value. 
There are also what
appear to be density `waves' propagating into the outer regions. As
the system relaxes, particles initially close to the center are pulled
into the potential well more tightly, causing the central density
enhancement.  Particles from farther out pass close to the center of
the distribution, gaining kinetic energy as they collapse.  They are
then  flung back to the outer regions and pass through incoming
particles. Finally they are turned around by the increasing central
attraction, which produces a `winding' of the DF.  Fig.\
(\ref{pdf_all}) shows how the collapse proceeds in phase-space for two
different collapse calculations.

The growth of the phase-mixing spiral `stream' in phase space is thus
perceived as a train of outward propagating density `waves' in
physical space. However they are really best understood in phase
space.  As the system is cooled, the peaks become much sharper,
eventually approaching the `smoothed infinite peaks' observed by
Fillmore \& Goldreich (1984) in their particle simulations.  

In addition to the $\alpha^2 = 2$ case considered above, simulations
were performed using very cold initial conditions ($\alpha^2 = 20$).
Qualitatively, many of the features seen above were also present in
this case (central density enhancement, density `waves' due to
multiple streaming).  One new feature, however, was observed in the
phase-space portrait.  An instability in the phase streams was seen
during the collapse (see Fig.\ (\ref{pdf_all}, bottom right panel) for
an example of the same instability as seen in the $n = 2$, $\alpha^2 =
10$ case).  This instability does not appear in warmer collapse
simulations (Fig.\ (\ref{pdf_all}), bottom left panel) when the
`smoothing' pre-exists.

This instability appears quite similar to one described by Henriksen
\& Widrow (1997).  In their shell-code simulations, the radial
instability was observed to blur the windings of the DF in phase space
and to produce a smoother, more relaxed DF (see their fig. 1).  Their
interpretation is consistent with our result in which the central
regions, still showing distinct streams after one characteristic time,
have at the resolution level of this simulation blended into a more
continuous DF after four characteristic times.  

It is conjectured that this instability is one of the mechanisms by
which a cold system can approach the smooth maximum-entropy state.
The observed instability spreads the infalling phase streams out and
allows an approach to a finer grained equilibrium.  The initially
warmer simulations are free of the instability due to the initial
spread in velocities. Reducing the discreteness of the velocity
streams appears to reduce the strength of the instability and the
initial DF is already sufficiently ``smeared'' in velocity space that
a smooth final state does not require the destabilizing of the phase
streams. The wave-particle aspect of `violent relaxation' must still
operate however.

 It was suggested in Henriksen \& Le Delliou (2002), that relaxation
 may be said to be complete when the finer grained (but still
 statistically valid) DF and less finely grained DF coincide. This is
 a useful notion but clearly it is bounded at the two ends of the
 resolution scale. If the resolution is such that an individual
 particle may be followed then clearly the DF description is not
 useful. On the other hand a resolution which just resolves the system
 would provide no structural information. There is then an optimum
 smoothing range over which to test the invariance of the relaxed
 DF. Practically, in our direct integration results, we simply coarse
 grain until until the DF is smoothly varying. This requires typically
 a $20$\% smoothing in velocity space. Our N-body calculations are
 presented unsmoothed.

\subsubsection{Velocity distribution}
\label{poly_vel_dist}
The extent (completeness) of any ultimate Gaussian will be limited by
the finite mass and  radius, since a complete Gaussian profile
(i.e.~extending to infinite positive and negative velocities) would
correspondingly require infinite mass and radius.  So, when we speak
of a Gaussian velocity distribution, it is with the implicit
understanding that it must be lowered in such a manner as to be
truncated at some finite velocity (since, of course, $f$ can never be
negative).

Lowering a Gaussian was considered by King (1966), who determined that
the cutoff and the method of lowering the curve have very little
effect on the central regions, with deviation in the density profile
appearing only near the spatial limit of the distribution.  A lowered
Gaussian will have the effect of including only those particles with
negative energy (bound particles). The method of modifying the DF to
include only those stars with negative energy has been discussed by
King (1966), Binney \& Tremaine (1987), and Kuijken \& Dubinski
(1994).  While it is true that a lowered Gaussian is not strictly a
Gaussian (as it does not lie within an infinite domain), up to the
point where $f = 0$ the velocity distribution should have a Gaussian
shape consistent with predictions (Nakamura,2000).

\begin{figure}
\plotone{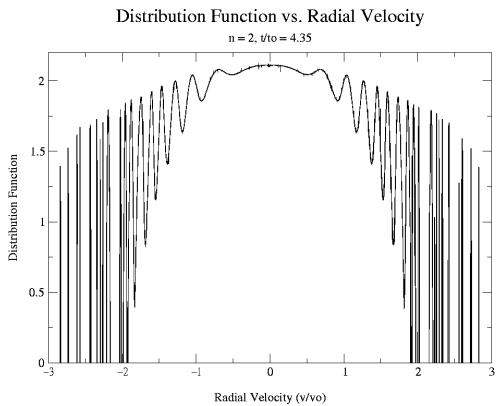}
\caption{\label{velocity_n2_a10_t435} 
Profile of DF vs.\ radial velocity for $n = 2$ polytrope with initial virial
ratio 0.1 after 4.35 characteristic times.  This profile is taken at $r
= 0$.} 
\end{figure}

\begin{figure}
\plotone{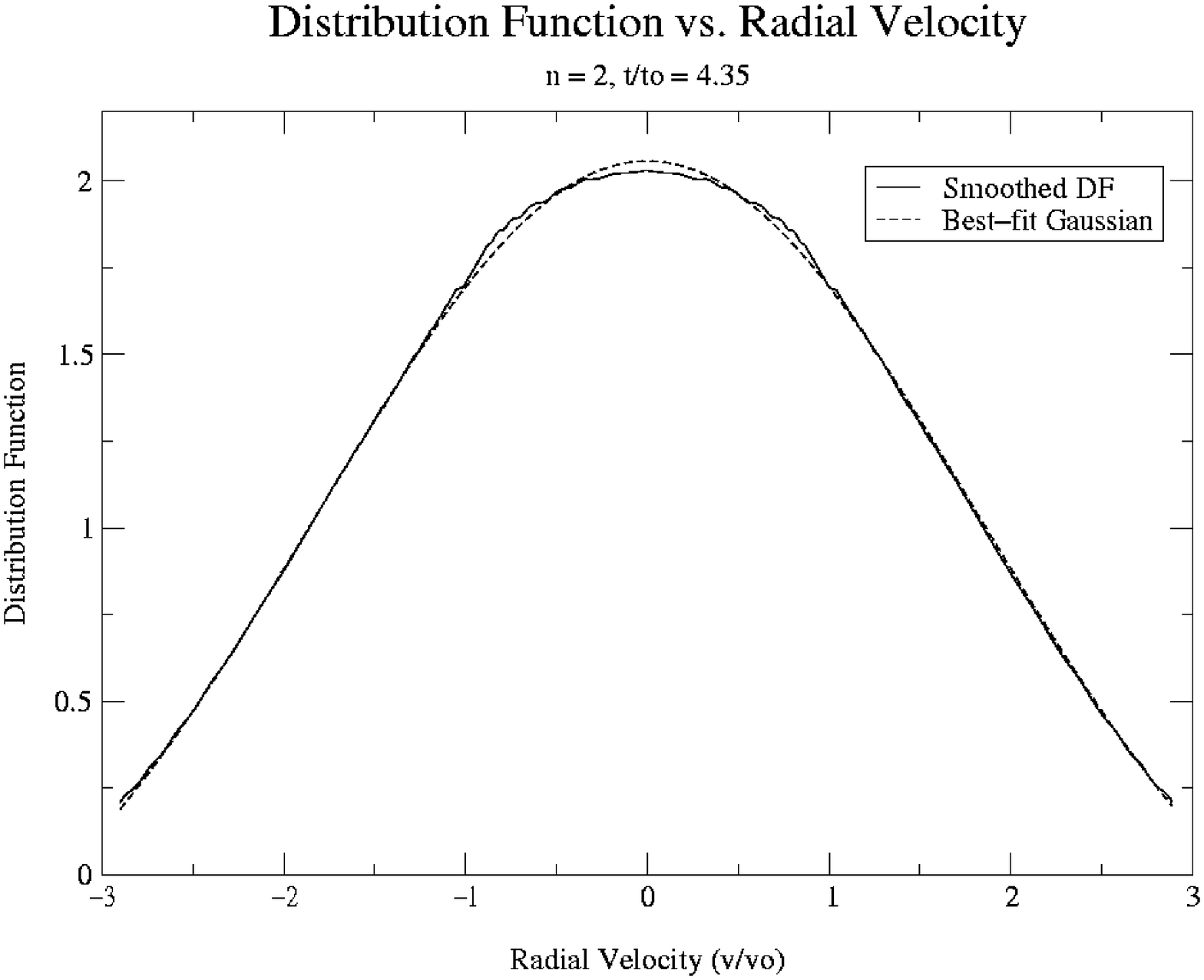}
\caption{\label{velocity_n2_a10_t435_s1000} 
Profile of DF vs.\ radial velocity for $n = 2$ polytrope with initial virial
ratio 0.1 after 4.35 characteristic times, smoothed with a window of
20\% of the total width.  Also shown is the best-fit Gaussian ($DF =
2.606 \;exp\!\left(-0.15 v_r^2\right) - 0.549$).  This profile is taken
at $r = 0$.}
\end{figure}

 As the destabilized polytrope collapses, the finest-grained DF spirals 
in phase space, leading to a series of peaks in the velocity distribution
(Fig.\ (\ref{velocity_n2_a10_t435})). 
By smoothing the velocity
distribution with a moving window average, we are able to determine a
coarser grained DF. We find that the smoothed DF is indeed Gaussian
(Fig.\  (\ref{velocity_n2_a10_t435_s1000})) up to the edge of the
distribution in velocity space. The figures are shown for the centres
of the polytropes.

The window function used in the smoothing was a simple top-hat
averaging, centered on the point under consideration.  A Gaussian
distribution has not been imposed anywhere in the calculation process, 
and the Gaussian shape of the smoothed distribution appears to be a
result of the relaxation process rather than an artifact of the
smoothing.  The width of the window was adjusted to smooth over all
the peaks in the finest-grained DF, while still maintaining a
significant `fine-grained' signal.  Choosing too small a window (too
high a resolution) does not sufficiently smooth the spikes (leaving
non-equilibrium features in the profile), while oversmoothing the
distribution with too large a window causes a suppression of the
signal. The signal is completely suppressed in the limit of the
coarsest-grained smoothing that would simply produce a completely 
flat DF. The top-hat width used was always the same fraction of the
total width of the velocity distribution in the various figures. 

It is clear that at the full resolution of this simulation the system
has not relaxed microscopically because the phase space `spiral' is
only beginning to be subject to the phase mixing instability. We would
expect that in time the smoothed or coarse-grained DF will become
valid on finer and finer scales. The calculation became very time
consuming at this point and so we decided to test the DF over longer time
scales with an N-body tree code. There (see below) one does see the
phase space structure becoming smooth at a fixed resolution.   

\begin{figure}
\plotone{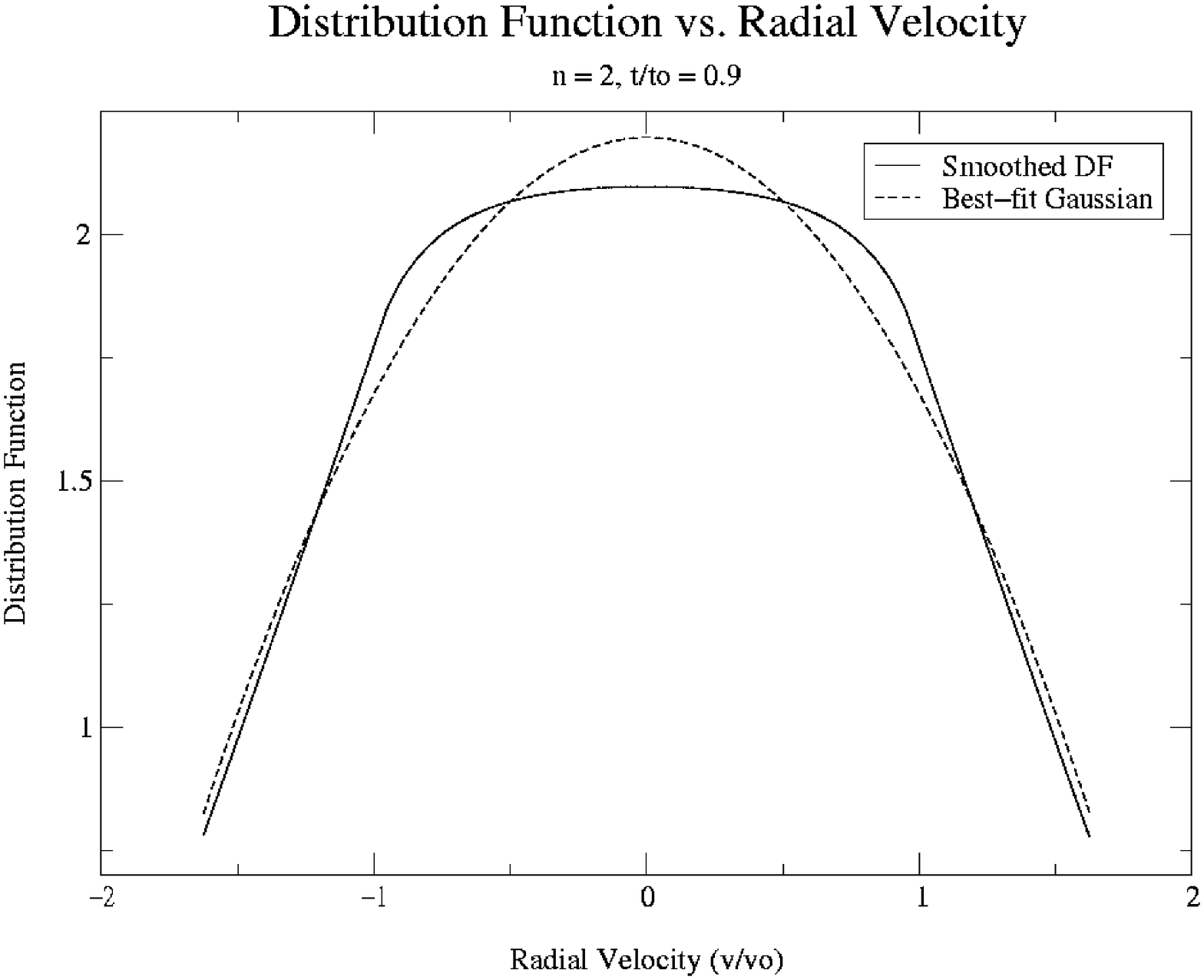}
\caption{\label{velocity_n2_a10_t090_s1000} 
Profile of DF vs.\ radial velocity for $n = 2$ polytrope with initial virial
ratio 0.1 after 0.9 characteristic times, smoothed with a window of
20\% of the total width.  Also shown is the best-fit Gaussian.  It is
clear that it this early stage in the relaxation process the DF has
not yet established a Gaussian velocity profile.  This profile is
taken at $r = 0$.}
\end{figure}

\begin{figure}
\plotone{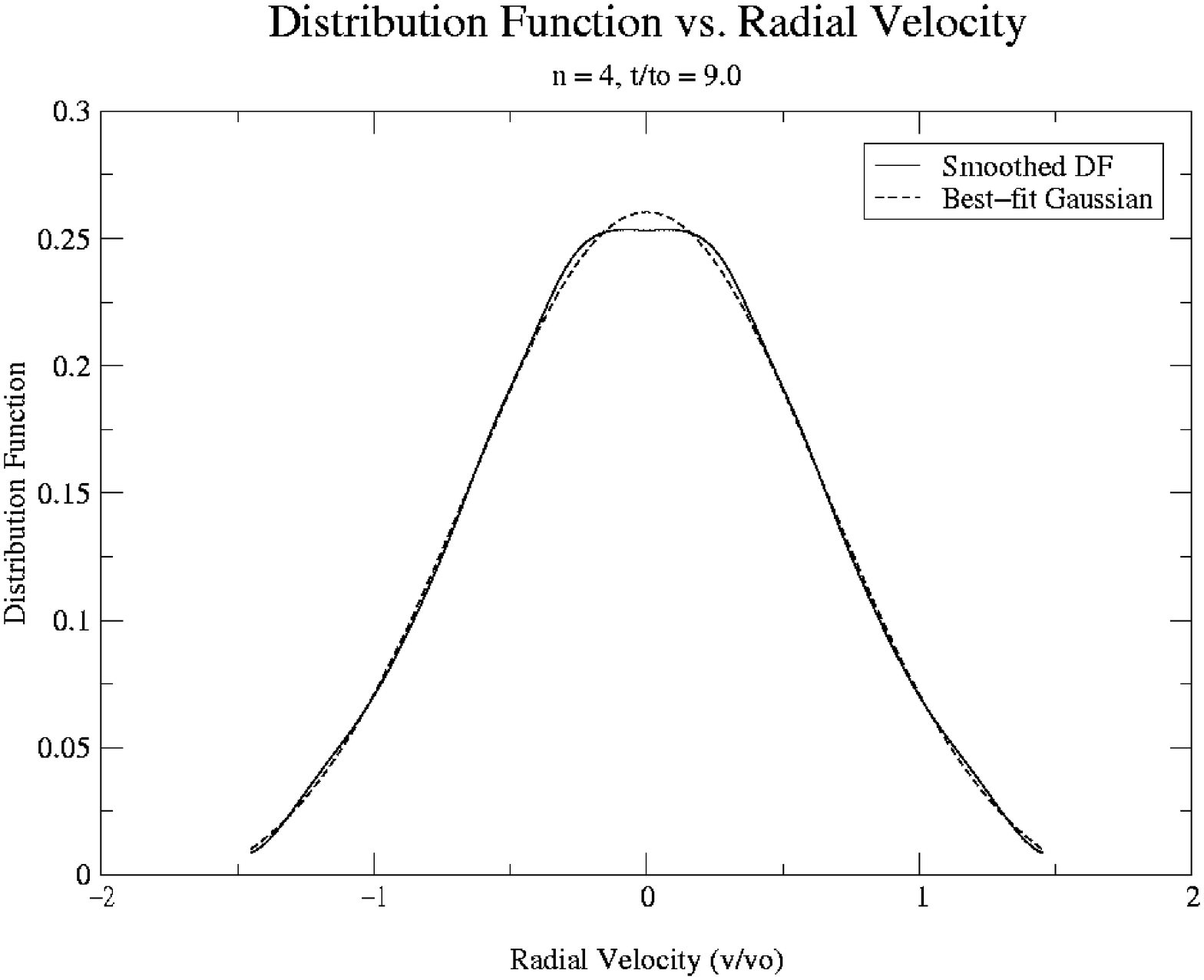}
\caption{\label{velocity_n4_a1.5_t900_s1000}
Profile of DF vs.\ radial velocity for $n = 4$ polytrope with initial virial
ratio 0.667 after 9.0 characteristic times, smoothed with a window of
20\% of the total width.  Also shown is the best-fit Gaussian.  This
profile is taken at $r = 0$.}
\end{figure}

A velocity distribution from an earlier time (before the central
region has approached its relaxed state) was smoothed with the same
window as Fig.\ (\ref{velocity_n2_a10_t435_s1000}), and is clearly
{\it not} Gaussian (Fig.\ (\ref{velocity_n2_a10_t090_s1000})).  
This
supports the idea that as time passes the distribution becomes more
nearly Gaussian, and that the Gaussian signals seen are not simply
artifacts of the smoothing.  Fig.\ (\ref{velocity_n4_a1.5_t900_s1000}) 
illustrates the effect for the $n = 4$ case.  This demonstrates that
the evolution to Gaussianity does not rely on a particular value of
$n$ to take place, and should occur for a general spherical collapse. 

The smoothing of the DF by a moving window average is an approximation
to a coarse-graining of the system.  In Henriksen \& Le Delliou (2002),
a coarse graining scheme is proposed which involves a nonuniform
rescaling of the phase space.  This scheme produces a series
representation of the DF which is interpreted as progressively finer
graining at higher orders.  They discover that the best behaved coarse
graining expansion is that which produces a Gaussian velocity
distribution in the inner (relaxed) region, in agreement with the
statistical arguments of Lynden-Bell (1967) and Nakamura (2000).  Our
current investigation confirms both the Gaussian inner distribution,
and their supposition that the deviation from the relaxed state
increases with radius.

\subsection{Collapsing Halo Results}
\label{halo_results1}
The input parameters for the KD code and the resulting dimensionless
halo properties are summarized in Tables
(\ref{halo_setup})--(\ref{halo_setup2}).  
\begin{deluxetable}{rlc|cll}
\tabletypesize{\scriptsize}
\tablecaption{\label{halo_setup} Kuijken \& Dubinski halo parameters
(Model 1)}
\tablewidth{0pt}
\startdata
$\Psi_{o}$    &= -8.0&&&Total Mass   &: 42.313$M_o$\\
$v_o$         &= $\sqrt{2}$&&&Tidal Radius &: 87.58$R_o$\\
$q$           &= 1&&&King Radius  &: 5.211$R_o$\\
$(r_c/r_k)^2$ &= 1&&&&\\
$R_a$         &= 1&&&&\\
\enddata
\end{deluxetable}
\begin{deluxetable}{rlc|cll}
\tabletypesize{\scriptsize}
\tablecaption{\label{halo_setup2} Kuijken \& Dubinski halo parameters
(Model 2)}
\tablewidth{0pt}
\startdata
$\Psi_{o}$    &= -8.0&&&Total Mass   &: 4.344$M_o$\\
$v_o$         &= $\sqrt{2}$&&&Tidal Radius &: 8.98$R_o$\\
$q$           &= 1  &&&King Radius  &: 0.5212$R_o$\\
$(r_c/r_k)^2$ &= 1  &&&&\\
$R_a$         &= 0.1&&&&\\
\enddata
\end{deluxetable}
In order to get physically
meaningful quantities from the scaled variables, we used the same
dark matter halo parameters as above (M$_{halo} \sim 1.3\times10^{12}
$M$_\odot$, R$_{halo} \sim \;160$ kpc) (Little \& Tremaine 1987;
Arnold 1992).  These values lead to the scalings presented in Tables
(\ref{halo_phys})--(\ref{halo_phys2}). 
\begin{deluxetable}{rl}
\tabletypesize{\scriptsize}
\tablecaption{\label{halo_phys} Kuijken \& Dubinski halo physical
scalings (Model 1)} 
\tablewidth{0pt}
\startdata
$M_o$                         &= 3.0723$\times10^{10}$ M$_{\odot}$\\
$R_o$                         &= 1.827 kpc\\
$t_o$ = $\sqrt{R_o^3/GM_o}$   &= 6.6425$\times10^6$ years\\
$v_o$ = $\sqrt{GM_o/R_o}$     &= 268.9 \kms\\
$\sigma_o$ = $v_o$/$\sqrt{2}$ &= 190.17 \kms\\
$T_{\mbox{core}}$ = $2\pi r_k/\sqrt{3}\sigma_o$ &= 18.90$t_o$\\
$T_{\mbox{cross}}$ = $2R_t/\sigma_o$ &= 175.16$t_o$\\
\enddata
\end{deluxetable}
\begin{deluxetable}{rl}
\tabletypesize{\scriptsize}
\tablecaption{\label{halo_phys2} Kuijken \& Dubinski halo physical
scalings (Model 2)} 
\tablewidth{0pt}
\startdata
$M_o$                         &= 2.9926$\times10^{11}$ M$_{\odot}$\\
$R_o$                         &= 17.817 kpc\\
$t_o$ = $\sqrt{R_o^3/GM_o}$   &= 64.79$\times10^6$ years\\
$v_o$ = $\sqrt{GM_o/R_o}$     &= 268.74 \kms\\
$\sigma_o$ = $v_o$/$\sqrt{2}$ &= 190.03 \kms\\
$T_{\mbox{core}}$ = $2\pi r_k/\sqrt{3}\sigma_o$ &= 2.67$t_o$\\
$T_{\mbox{cross}}$ = $2R_t/\sigma_o$ &= 25.41$t_o$\\
\enddata
\end{deluxetable}

The N-body haloes produced by the KD94 code were destabilized by
reducing the velocities by a constant factor just as in the polytropic
collapse considered above.  This reduced the thermal support and
facilitated the collapse of the halo.  In the first case, we used a
velocity reduction factor of $\alpha^2 = 2$ applied to a Model 1 halo.
This is a fairly warm collapse and, as such, we do not expect to see
significant growth of the Henriksen \& Widrow (1997) phase mixing 
instability.

\begin{figure}
\plotone{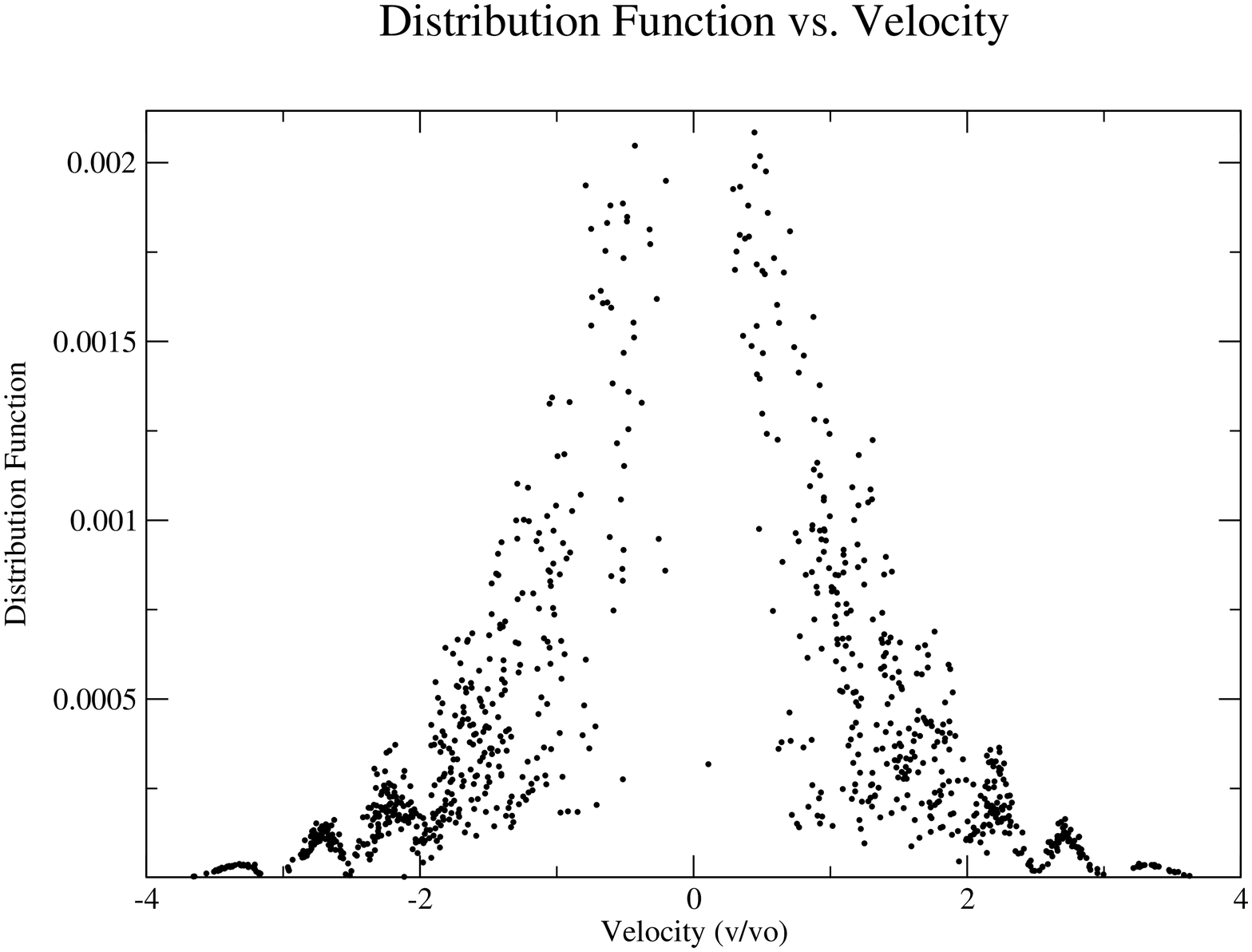}
\caption{\label{halo_vel_2.0_peak}
Profile of DF vs.\ radial velocity for a galactic halo (Model 1) with
initial virial ratio 0.5, prior to relaxation of the halo ($t/t_o =
20$).  This slice is taken at a radial bin of width 0.05, centered at
2.0.  Peaks in the velocity distribution corresponding to distinct
phase streams are apparent in the wings of the distribution.}
\end{figure}

The relaxation of the Model 1 halo proceeds in much the same manner as for
the spherical polytropes calculated with the CBE code above.  As the
collapse progresses, the velocity distribution spreads in order to
provide the necessary thermal support to stabilize the configuration.
The peaks shown in Fig.\ (\ref{halo_vel_2.0_peak}) were similar to
those seen in Fig.\ (\ref{velocity_n2_a10_t435}) and were evident in
the collapse at various radii. 
The finite resolution coming from the
discrete nature of the N-body simulation causes the peaks in the
velocity distribution to become indistinct and blur in time into a
more continuous distribution.

\begin{figure}
\plotone{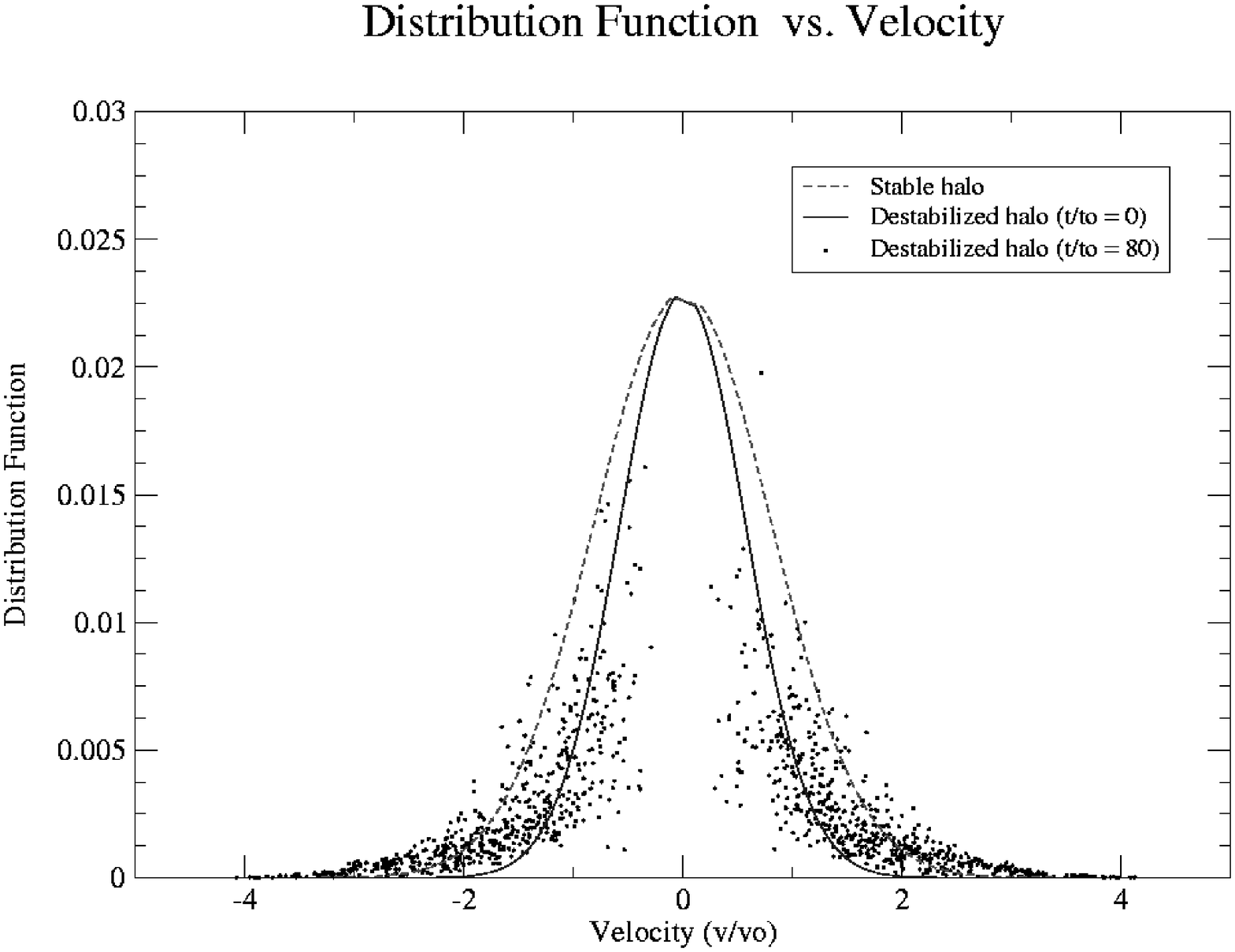}
\caption{\label{halo_vel_0.9}
Profile of DF vs.\ radial velocity for a galactic halo (Model 1) with
initial virial ratio 0.5.  This slice is taken at a radial bin of
width 0.05, centered at 0.95.}
\end{figure}

\begin{figure}
\plotone{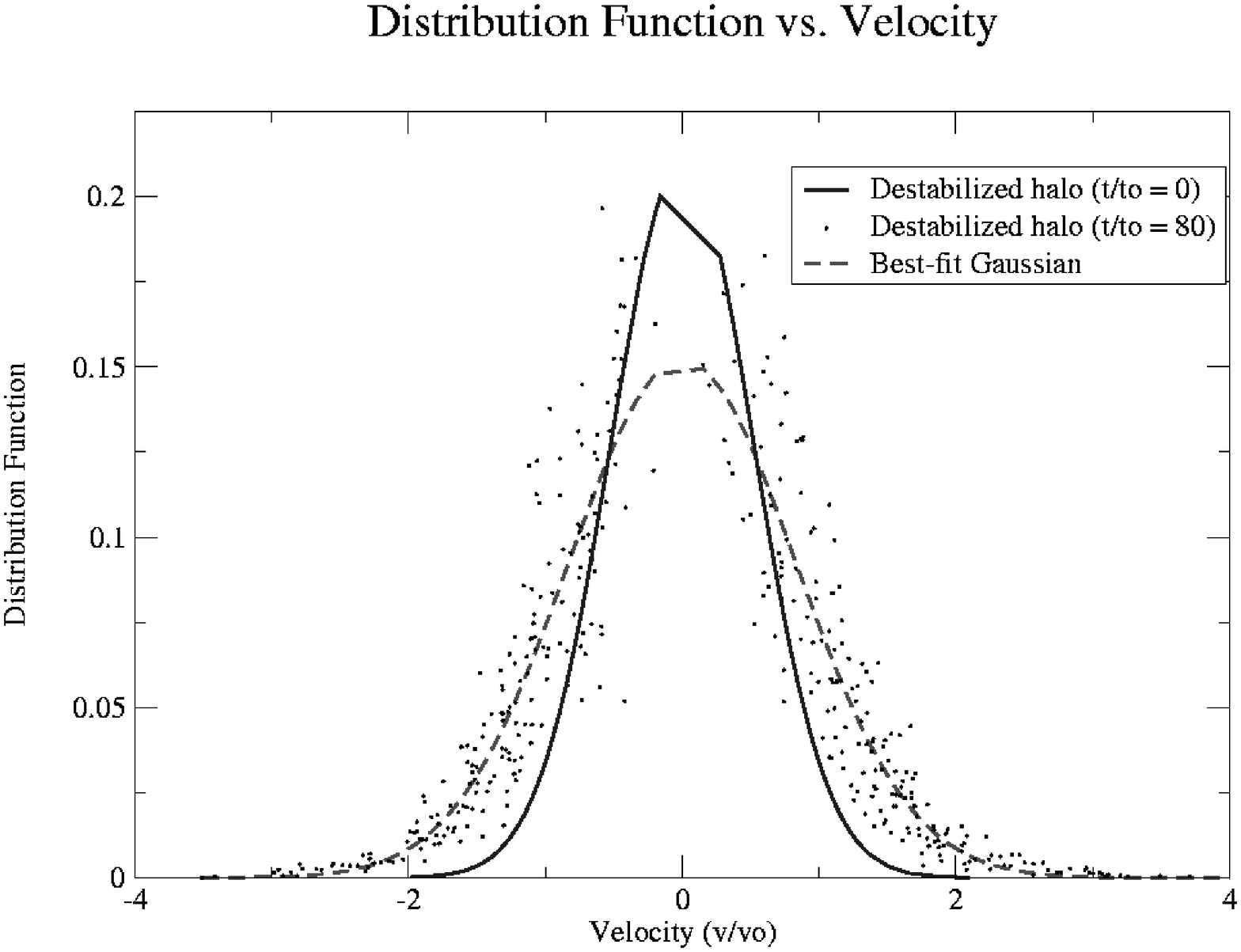}
\caption{\label{halo_vel_0.15}
Profile of DF vs.\ radial velocity for a galactic halo (Model 1) with
initial virial ratio 0.5.  This slice is taken at a radial bin of
width 0.05, centered at 0.15.  The flattening at the peak of the
initial and stable halo distributions is simply a result of having too
few points with near-zero velocity to completely define the curve.}
\end{figure}

Fig.\ (\ref{halo_vel_0.9}) shows the increasing velocity width in the
approach to the equilibrium state at $r=0.9$.   Fig.\
(\ref{halo_vel_0.15}) shows the velocity distribution closer to the
center of the halo ($r=0.15$, as compared to $r=0.9$).  At this time,
the velocity distribution in this central region has already evolved
to a near-Gaussian form (Fig.\ (\ref{halo_vel_0.15})), but the
distribution farther out still shows evidence of non-Gaussianity in
the form of noticeable streams in the wings of the distribution (Fig.\
(\ref{halo_vel_2.0_peak}) -- (\ref{halo_vel_0.9})).  In time the
velocity distributions at larger radii will also evolve to
near-Gaussian form as they spread to provide `thermal' support.

For the Model 2 halo, an initial virial ratio of 0.25 was chosen.
This value is on the edge of the `cool' domain where we expect the
Henriksen \& Widrow (1997) instability to play a r\^{o}le in
relaxation.
\begin{figure*}
\begin{tabular}{cc}
\rotatebox{0}{\scalebox{0.30}
        {\includegraphics
                {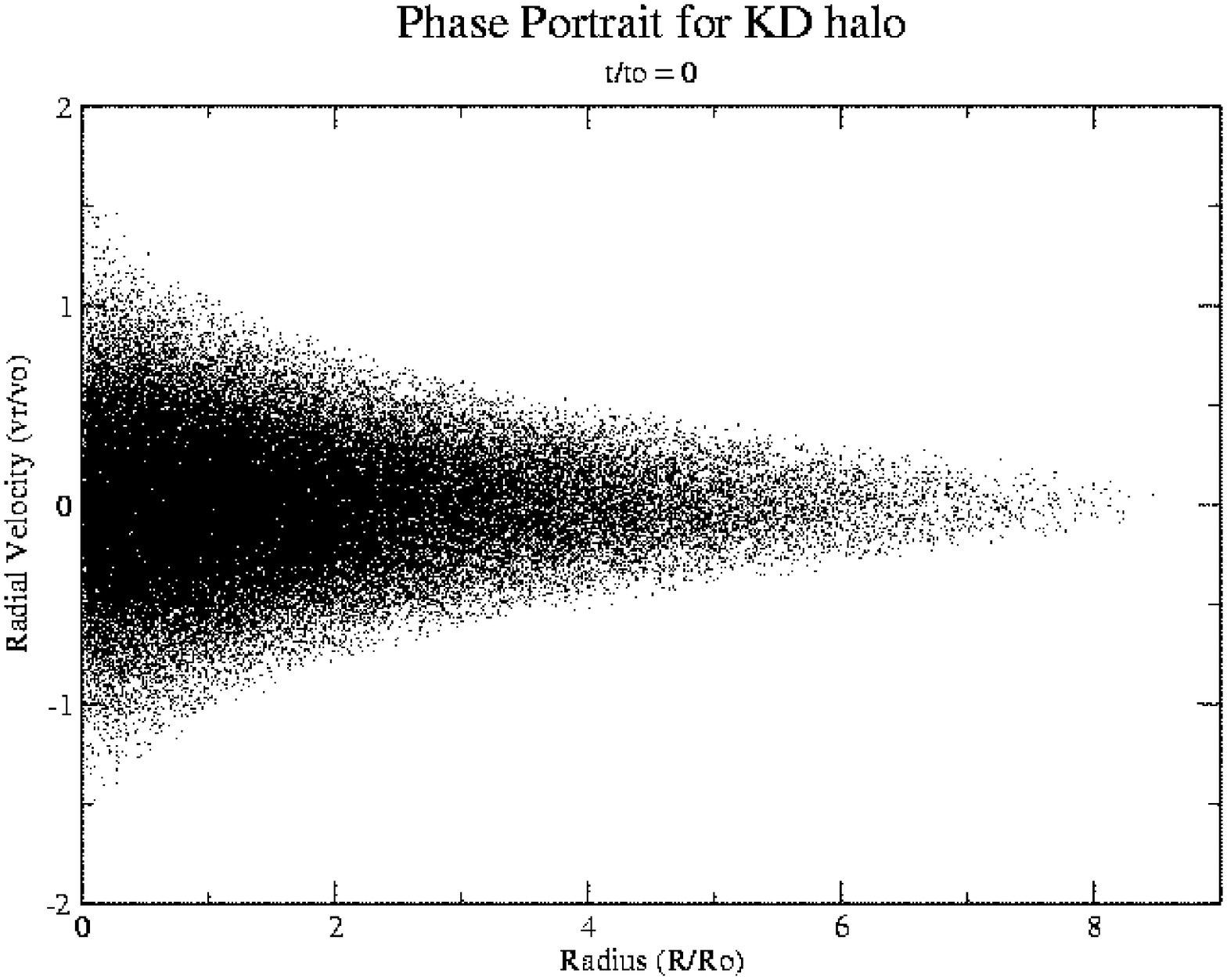}}} & 
\rotatebox{0}{\scalebox{0.30}
        {\includegraphics
                {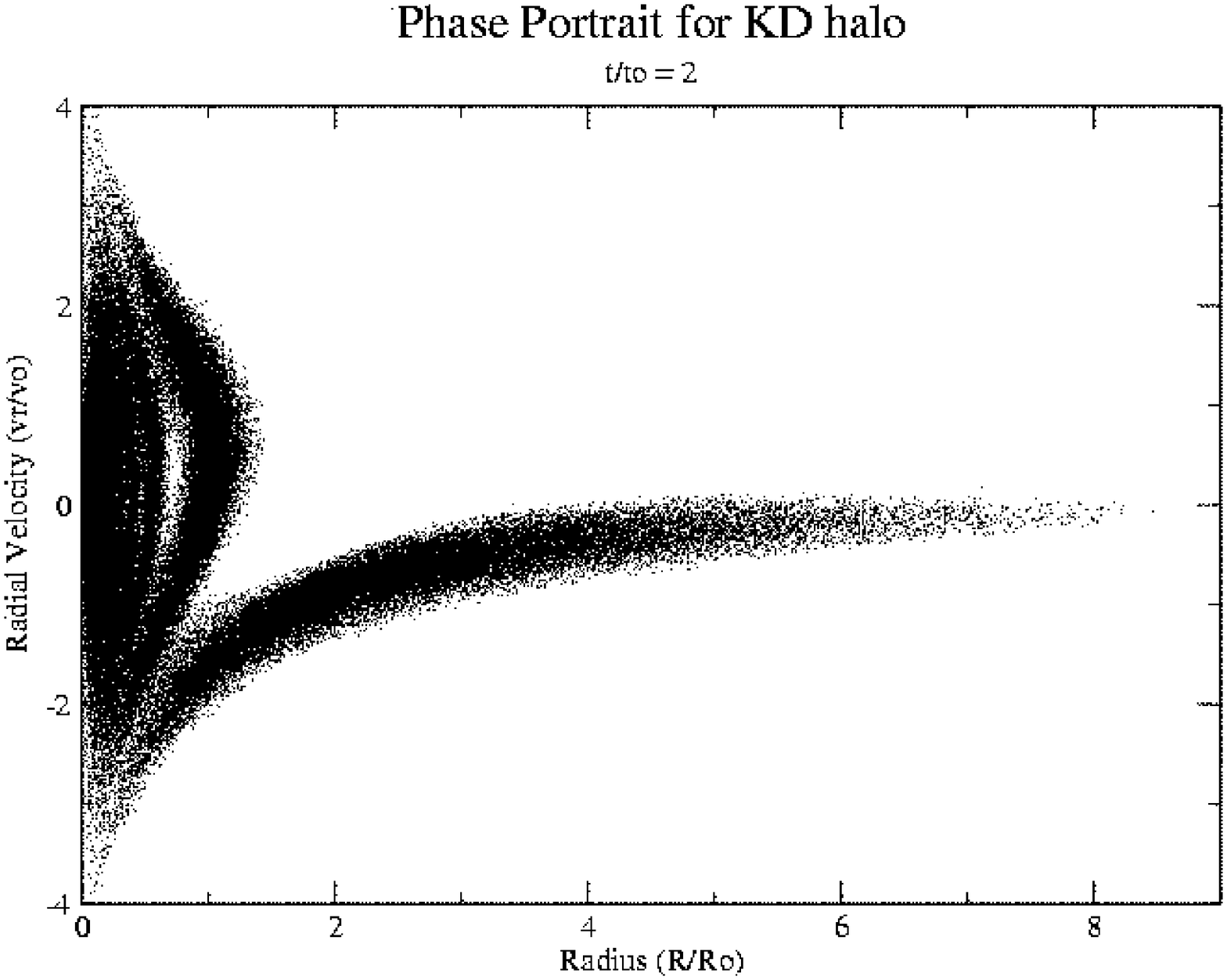}}} \\
\rotatebox{0}{\scalebox{0.30}
        {\includegraphics
                {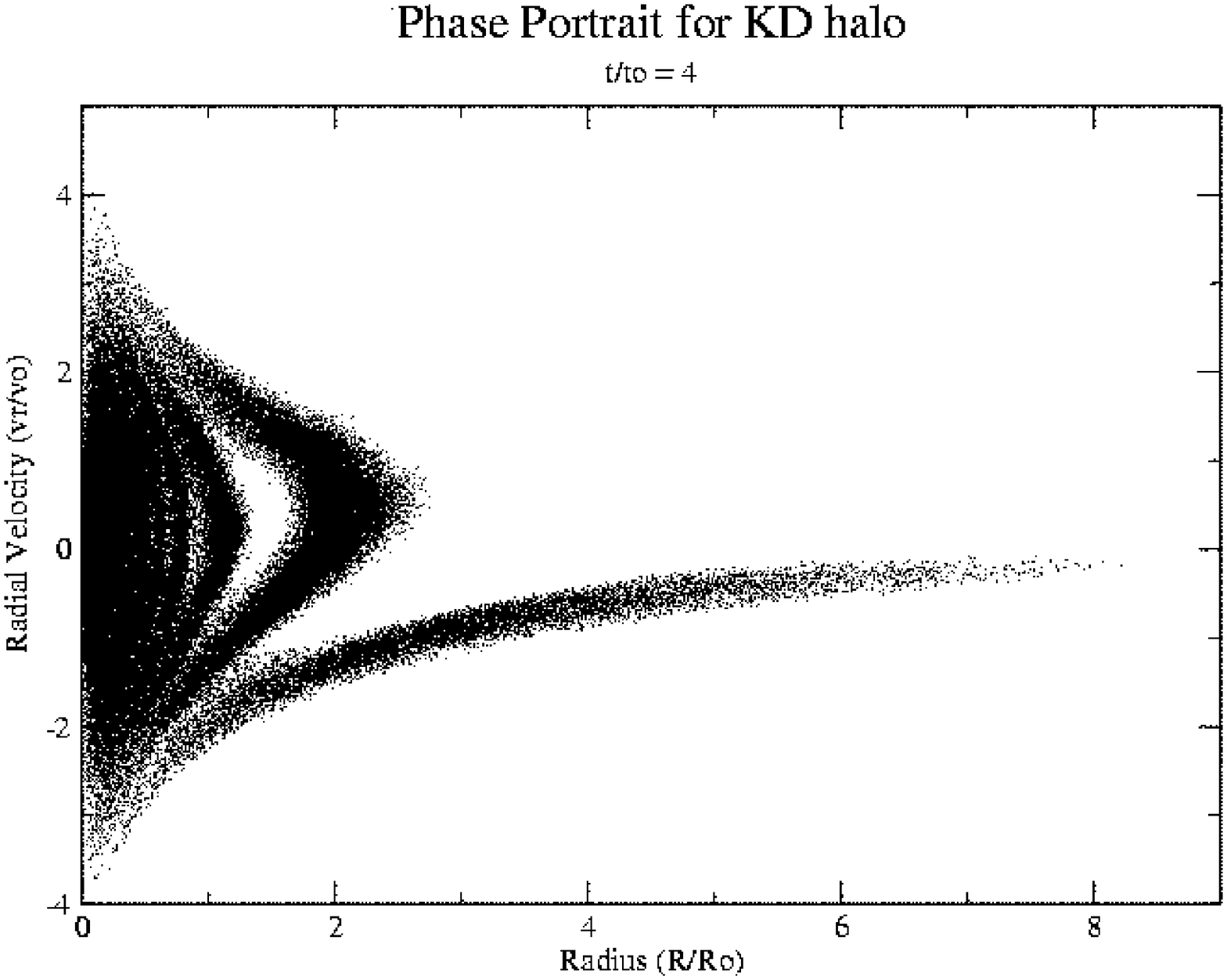}}} & 
\rotatebox{0}{\scalebox{0.30}
        {\includegraphics
                {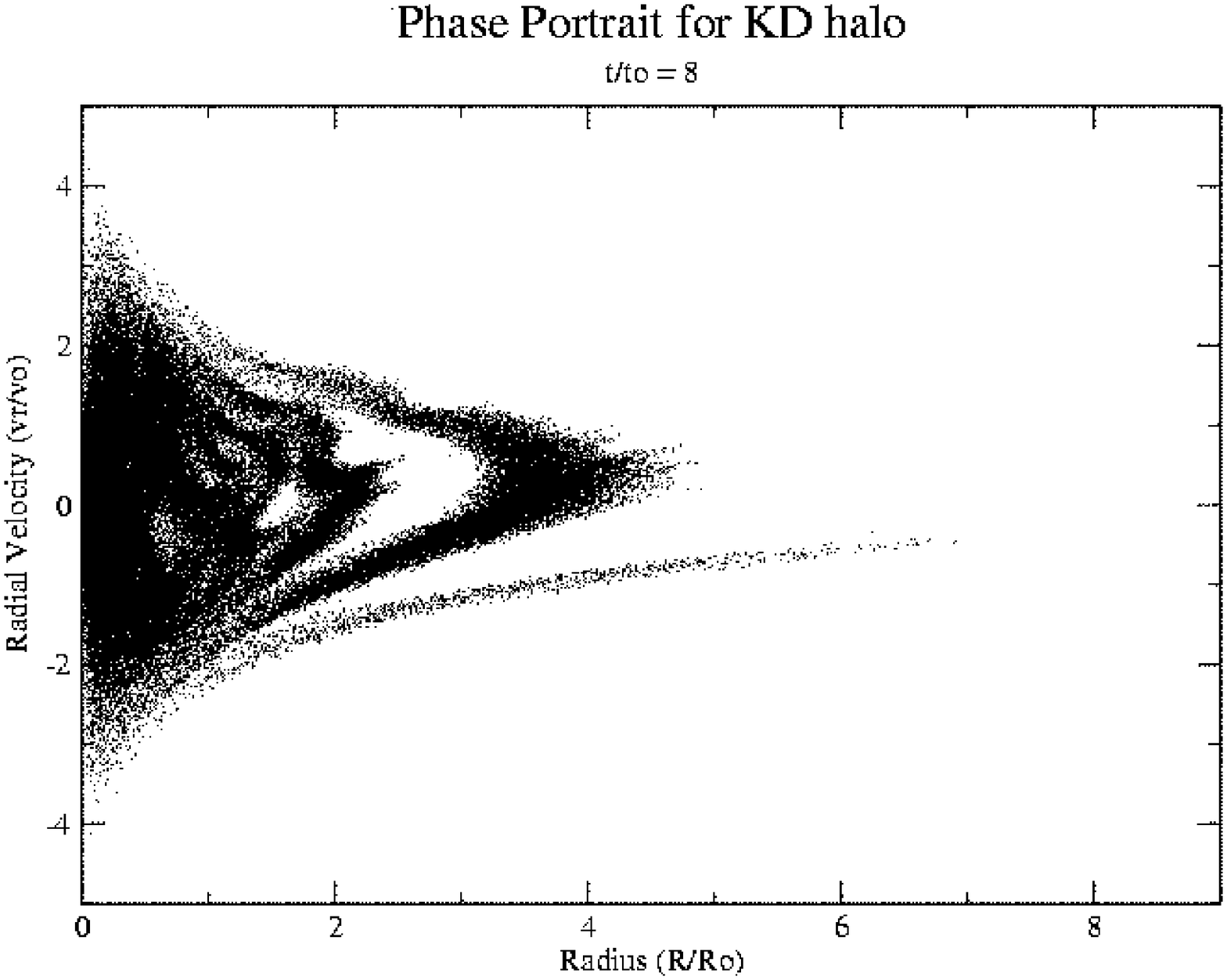}}} \\
\end{tabular}
\caption{\label{phase_vel_4.0}
($r$, $v_r$) projection of a collapsing KD halo (Model 2) with
$\alpha^2 = 4$.  Time increases from left to right, top to bottom.
Times shown are $t/t_o$ = 0, 2, 4, 8.}
\end{figure*}

\begin{figure}
\plotone{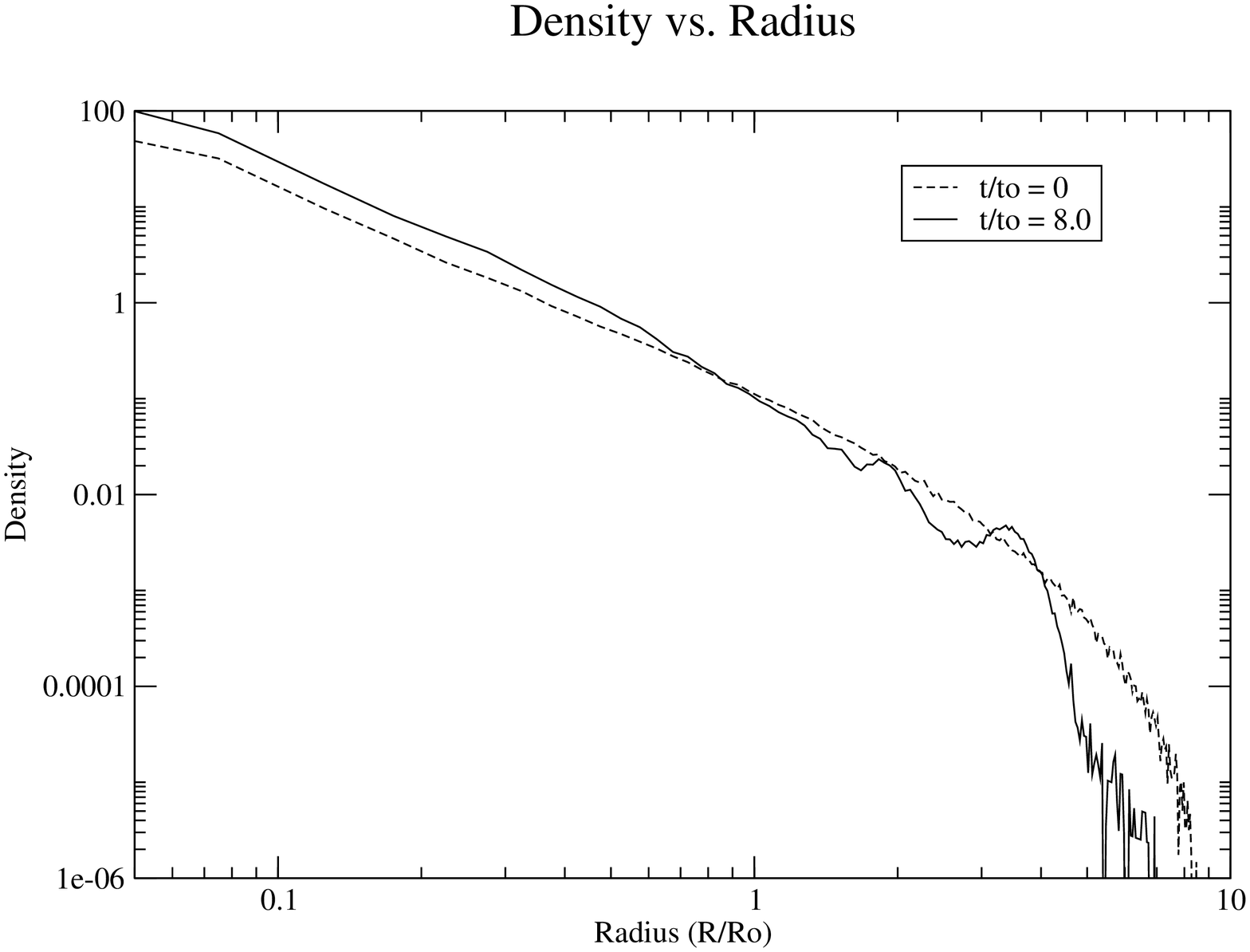}
\caption{\label{model2_density} 
Evolution with time of density for a galactic halo (Model 2) with
initial virial ratio 0.25.  Density waves are clearly seen propagating 
outward.  The power-law region of the final state goes as $r^{-3}$.} 
\end{figure}

\begin{figure}
\plotone{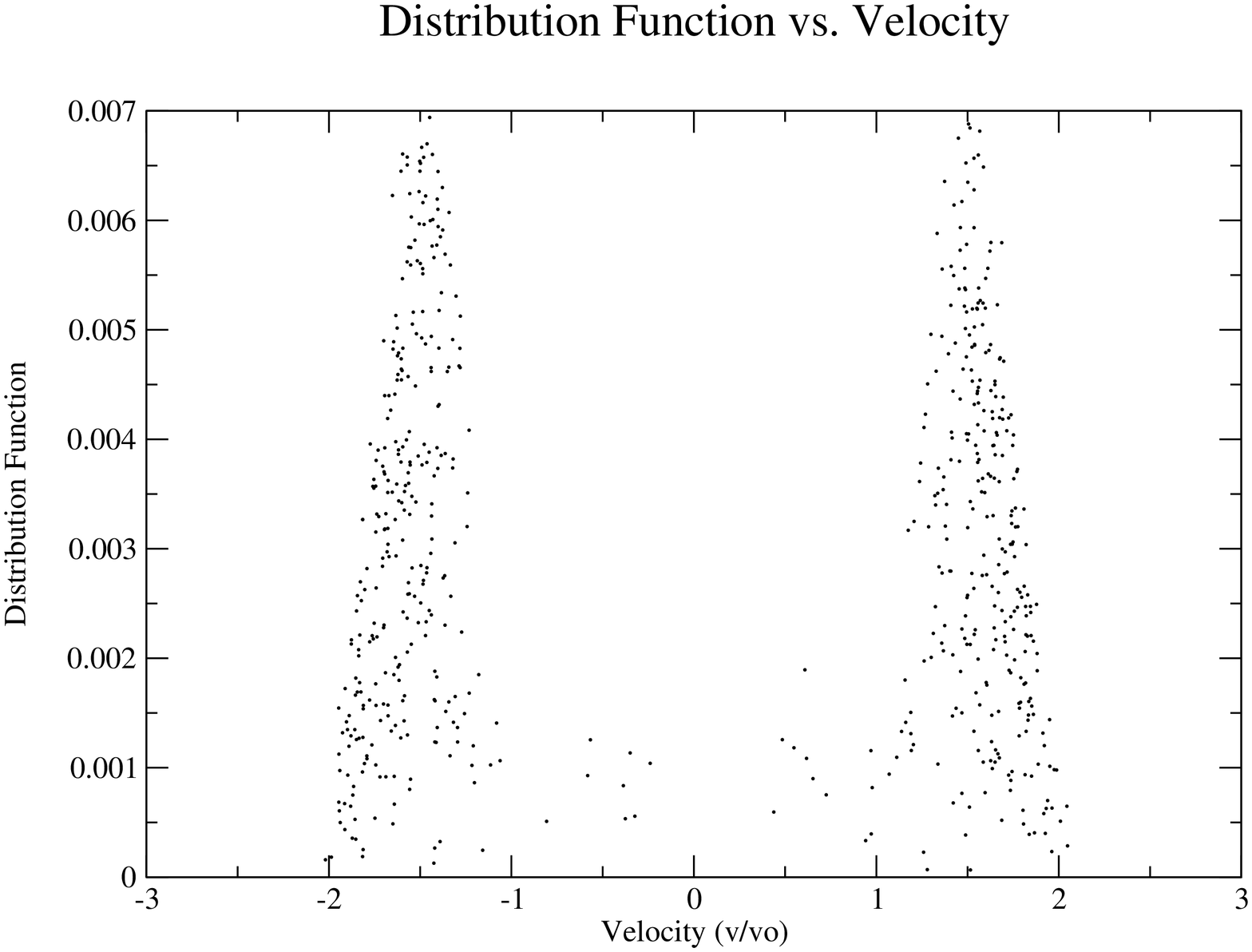}
\caption{\label{halo_vel_4.0_peak1}
Profile of DF vs.\ velocity for a galactic halo (Model 2) with
initial virial ratio 0.25.  This slice is taken at a radial bin of
width 0.05, centered at 0.75.  The velocity distribution is dominated
by two large peaks (velocity streams).}
\end{figure}

\begin{figure}
\plotone{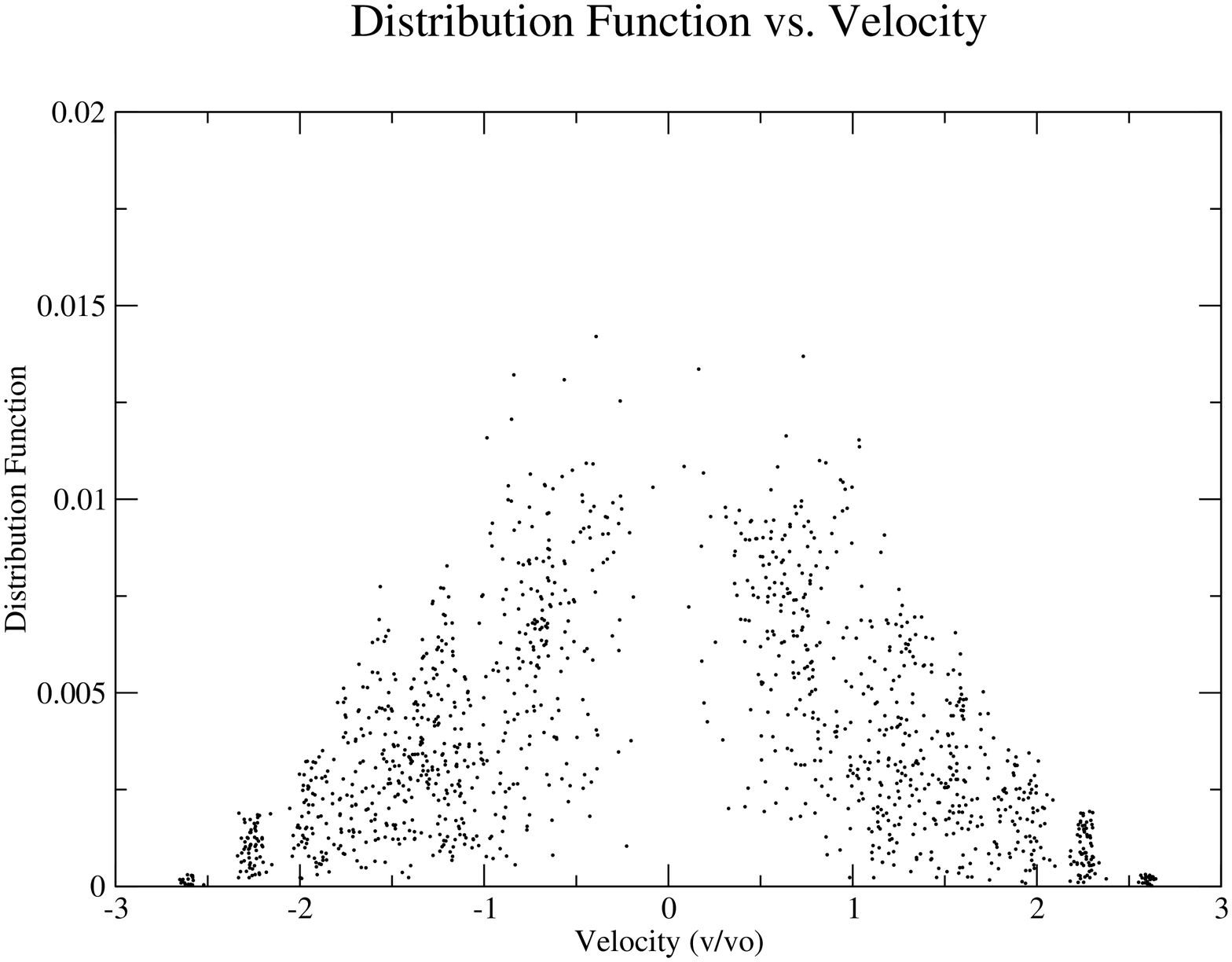}
\caption{\label{halo_vel_4.0_peak2}
Profile of DF vs.\ velocity for a galactic halo (Model 2) with
initial virial ratio 0.25.  This slice is taken at a radial bin of
width 0.05, centered at 0.75.  The velocity distribution is in the
process of relaxing to Gaussianity, and the peaks in the wings of the
distribution which dominated the velocity distribution in Fig.\
(\ref{halo_vel_4.0_peak1}) have become much less prominent due to the
onset of the Henriksen \& Widrow \cite{hw97} instability.}
\end{figure}

\begin{figure}
\plotone{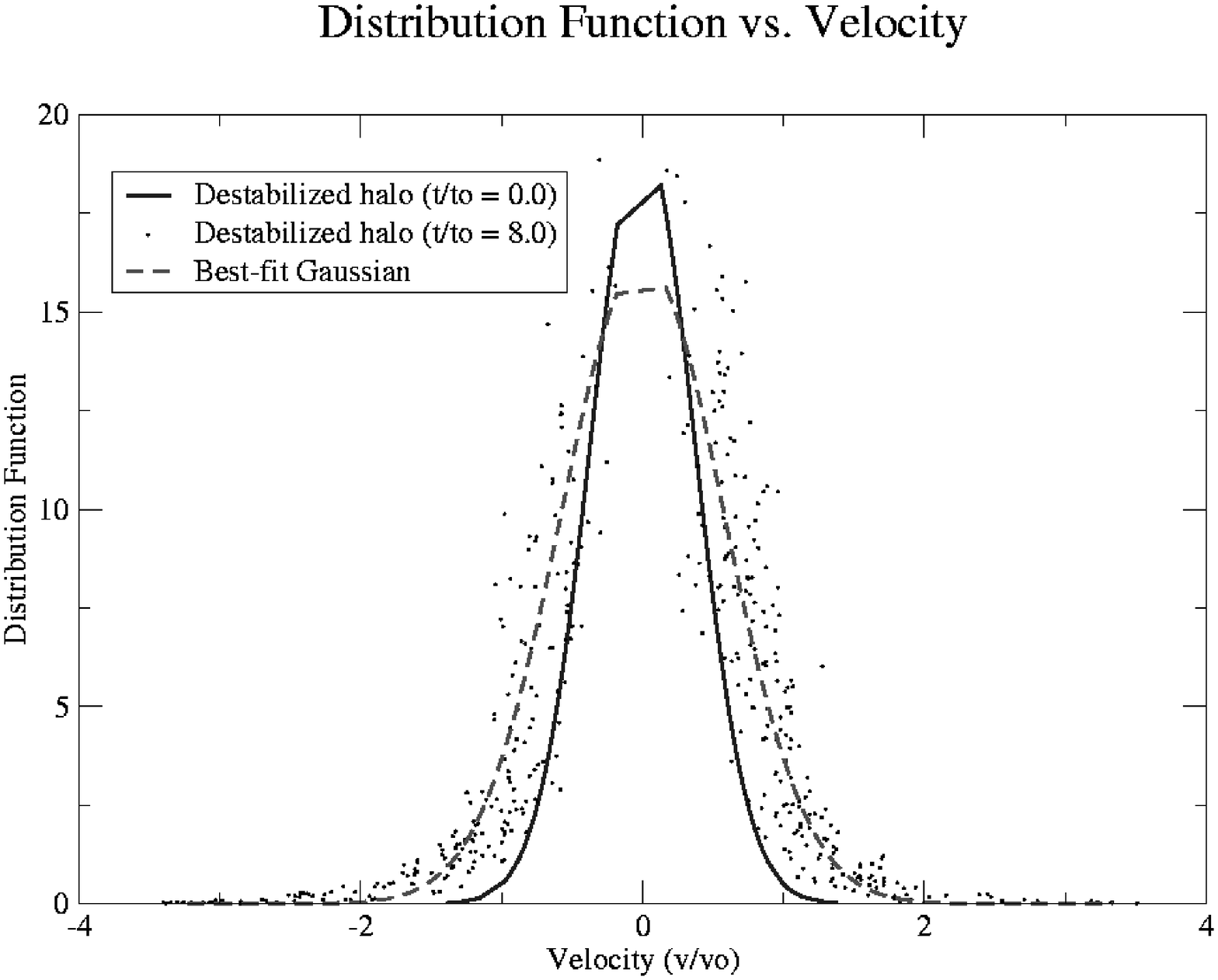}
\caption{\label{halo_vel_0}
Profile of DF vs.\ velocity for a galactic halo (Model 2) with
initial virial ratio 0.25.  This slice is taken at a radial bin of
width 0.05, centered at 0.025.  The velocity distribution has almost
reached a relaxed Gaussian state.}
\end{figure}

As the collapse proceeds, we find very similar results to
those found with the CBE calculation.  The DF phase mixes in the
projected ($r$, $v_r$) phase-space (Fig.\ (\ref{phase_vel_4.0})) --
also observed in configuration space as a set of outward propagating
density `waves' (Fig.\ (\ref{model2_density})) 
and again in the
velocity distribution as a set of peaks (seen dominating the velocity
distribution in Fig.\ (\ref{halo_vel_4.0_peak1})), 
which spread and
become less distinct  in time (Fig.\ (\ref{halo_vel_4.0_peak2})).
The onset of the phase mixing instability can be seen in the bottom
right panel of Fig.\ (\ref{phase_vel_4.0}).  We note that the central
regions have already become close to Gaussian (Fig.\
(\ref{halo_vel_0})).  
The width of the distribution has once again
increased to provide the thermal support required to halt collapse.

\begin{figure}
\plotone{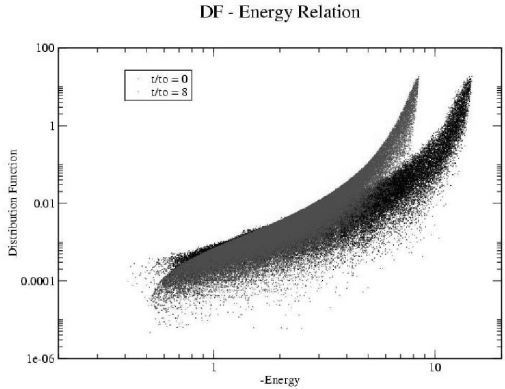}
\caption{\label{halo2_DF_E}
Distribution Function -- Energy relationship for a KD halo (Model 2)
with an initial virial ratio of 0.25.  For this halo, the rough shape
of the relation is maintained, with the most negative energy particles
becoming more tightly bound as collapse progresses.}
\end{figure}

The DF--Energy correlation is plotted in Fig.\ (\ref{halo2_DF_E}).
As the halo relaxes to its new equilibrium, we can see that the
overall shape of the curve is maintained, although the larger phase space
density (higher $f$) particles (corresponding to those that initially
have more negative energy) becoming even more tightly bound as time
passes.  Funato, Makino \& Ebisuzaki (1992a) also observed this energy
segregation through violent relaxation, but they interpreted this as
meaning that the Gaussian distribution would not be realized in
practice.  Our results demonstrate that although violent relaxation
does segregate energies, a Gaussian velocity distribution can
nevertheless result from the collapse.

The fact that the shape of the $f(E)$ relation is maintained suggests that
the particles retain some memory of their initial state throughout the
relaxation process.  This is consistent with the results of van Albada
(1982), who also observed correlation between initial and final
energies in violently-relaxing particles as did Henriksen and Widrow
(1999).  This `moderate' Violent relaxation is nevertheless a
process which acts over a much shorter timescale than two-body
relaxation in the systems studied.  Thus we conclude that a
collisionless collapse needs only a moderate dispersion in energy at
each position,in order to produce a Gaussian velocity distribution.

The behaviour of the virial ratio is shown for a typical case in Fig.\
(\ref{vr_many}). We see not only the finite number effects referred to
previously but the trend toward an ultimate virial ratio different
from unity. This is a numerical effect which results from calculating
the potential energy using the exact summation over the $1/r$
potentials of a set of pointlike particles in a system which evolves
under a softened force.  This effect is well known and is due to the
slight inconsistency of evolving the system under a softened
gravitational force, while calculating the potential energy from the
unsoftened potential. 

\subsection{Colliding Halo Results}
\label{halo_results2}
In this section spherical symmetry is explicitly broken by modelling
the merging of two centrally cusped galactic haloes.  We performed six
merger simulations.  All six consisted of the merging of two identical
stable galaxy models (KD Model 2 -- see Tables \ref{halo_setup2} \&
\ref{halo_phys2} for halo parameters and physical scalings
respectively).  The collision parameters are summarized in Table
\ref{collision_parameters}.

In the first simulation, the two galaxies were placed with their
centers separated by a distance $54 R_o = 0.962$ Mpc, with a relative
approach velocity of $1.4 v_o = 376.24$\kms.  This approach velocity
is approximately half of what it would be if the galaxies had come
from infinity with zero initial velocity.  In the second simulation,
the galaxies  were initially at rest with respect to one another with
their centers separated by $24 R_o = 427.6$ kpc. The third simulation
had the galaxies starting from a separation of almost $2.5$ Mpc, with
initial velocities of $40.3$\kms.  The initial conditions of this
third simulation are such that the difference in the interaction
(tidal) gravitational potential across each galaxy is very small. This
was intended to minimize the initial disequilibrium of the haloes.
The fourth simulation began with the galaxies already overlapping and
with zero relative velocity (more like a fragmentation).  The centers 
are separated by $4.5 R_o = 80$ kpc (which places the edges of the
King radii approximately $4.5 R_o$ apart).  The fifth and sixth
``collision'' (again more like fragmentation) simulations  had the
cores of the haloes initially overlapping, and zero relative velocity.
Simulation 5 begins with zero distance between the centers of the
haloes, and simulation 6 begins with the centers separated by $1 R_o =
17.8$ kpc.  The initial distance between the haloes in simulation 6
places the King radii of the haloes in tangential contact.

The virial ratio and total energy are both well conserved in these
simulations, with the virial ratio approaching unity (but see comment
in the previous section) shortly after the merging and the energy
remaining within $1\%$ of the initial value throughout the lifetime of
the simulation.  The simulations were allowed to run for almost 12 Gyr
(simulation 1), over 17 Gyr (simulation 2), over 50 Gyr (simulation
3), 13 Gyr (simulation 4), and 1.3 Gyr (simulations 5 \& 6). 

\begin{figure}
\plotone{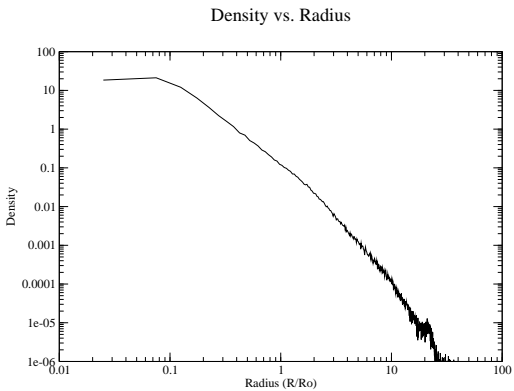}
\caption{\label{collide_rho} 
Final density profile for a pair of colliding galactic haloes
(Simulation 2).  This figure is taken at a time $t=10.1$ Gyr after the
initial contact.  The power-law region of the final state goes as
$r^{-3}$ between the edge of the core region and the first outgoing
density peak at $\sim 25 R_o$.} 
\end{figure}

The final density profile of simulation 2 is typical of that found in
the first three merger simulations, and is shown in Fig.\
(\ref{collide_rho}).  
It is seen to exhibit a flat central region,
with a power-law envelope which falls off as $r^{-3}$.  This
observation is consistent with the results of White \cite{white78}
\cite{white79}, who found the same behaviour in an extensive
examination of elliptical galaxy mergers.

\begin{figure}
\plotone{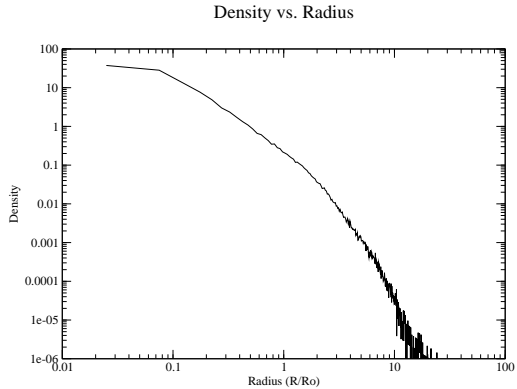}
\caption{\label{collide_rho4} 
Final density profile for a pair of initially-overlapping galactic
haloes (Simulation 4).  This figure is taken after $t=12.9$ Gyr of
relaxation.  The inner power-law region goes as $r^{-2.2}$, and the
outer power-law region goes as $r^{-4.25}$.} 
\end{figure}

Simulations 4 -- 6 have final density profiles which are different
from $r^{-3}$ and, in fact, show two different power-law regions in
addition to the flat central core. It is interesting that this
difference should characterize a kind of re-merging after an earlier 
fragmentation rather than the violent collisions of the first three
mergers. The effect is to suppress the complete transition to $r^{-3}$ --
leading to an inner region which is flatter than $r^{-3}$, and an
outer region which is steeper.  The final density profile of
simulation 4 is shown in Fig.\ (\ref{collide_rho4}). 
This is much like
the NFW profile {\it except that there is a flat core}. 

\begin{figure}
\plotone{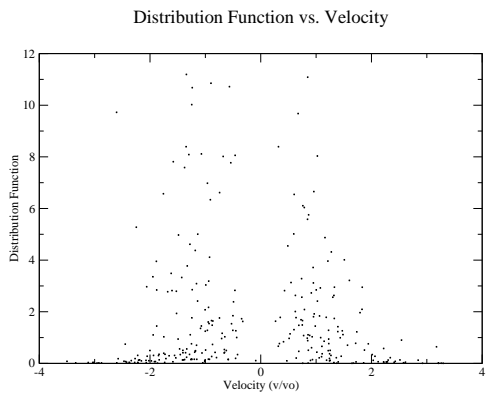}
\caption{\label{collide_vel1}
Profile of DF vs.\ velocity for colliding galactic haloes (Simulation
1).  This slice is taken at a radial bin of width 0.05, centered at
0.025.  The velocity distribution appears centrally peaked, but is not
clearly Gaussian.}
\end{figure}

\begin{figure}
\plotone{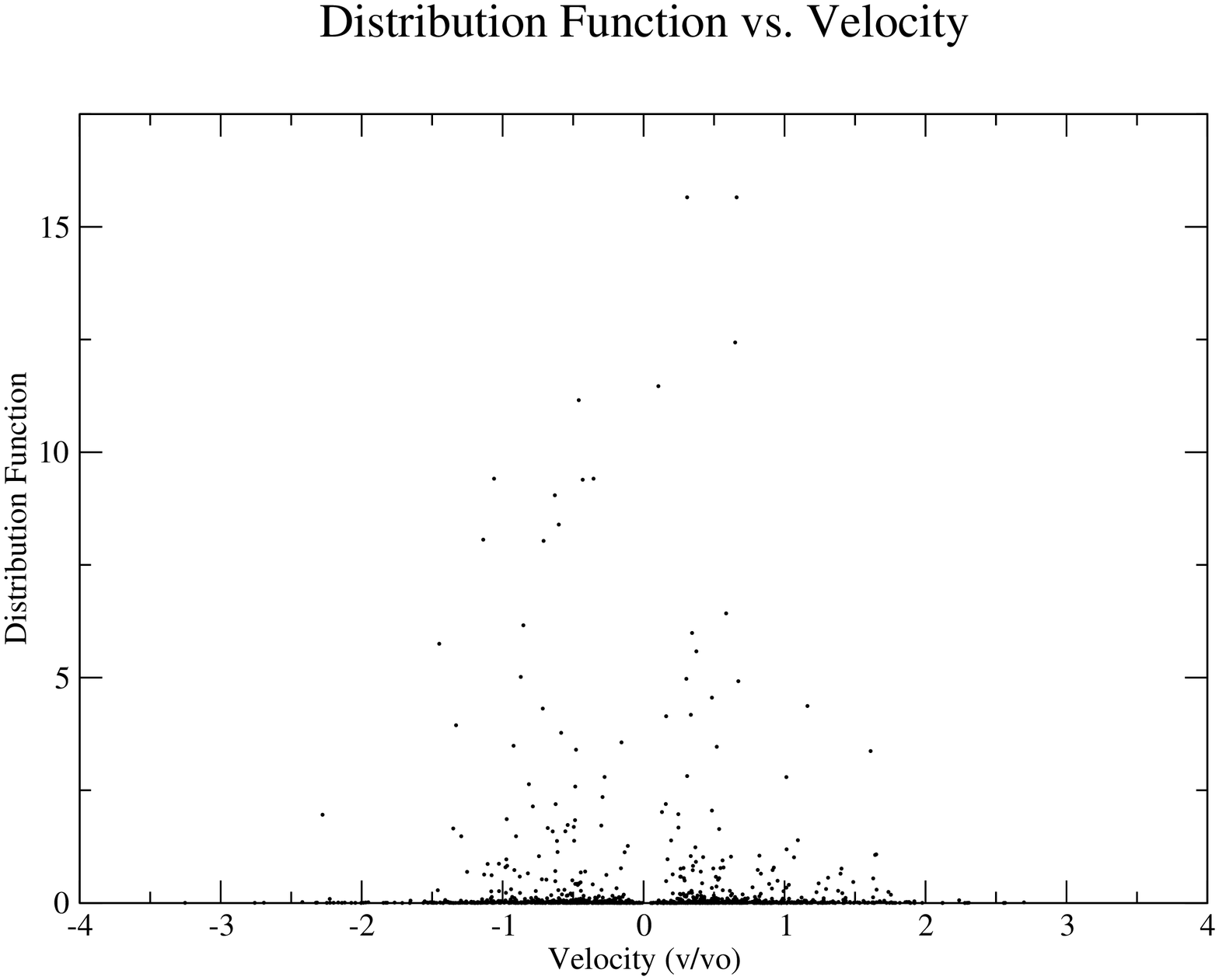}
\caption{\label{halo.collide3.vel.j0}
Profile of DF vs.\ velocity for colliding galactic haloes (Simulation
3). This figure is taken at an angular momentum slice $0 \leq j^2 <
0.1$. The velocity distribution appears centrally peaked, but is not
clearly Gaussian.}
\end{figure}

As all of the haloes relax, the velocity distributions for the
collision simulations become centrally peaked at all radii.  However,
it is not possible to state with any degree of confidence that they
are Gaussian (Fig.\ (\ref{collide_vel1})), at least for simulations 1,
2, and 3.  
Examining the velocity distribution for a small range of
$j^2$ (rather than for all values of the angular momentum), and
angular momentum direction ($\cos{\theta} = j_z/\sqrt{j^2}$) does not
improve the Gaussian fit.  An example is given in in Fig.\
(\ref{halo.collide3.vel.j0}).  
This indicates that the velocity
distribution is not segregated by angular momentum with distinct
Gaussians appearing in individual $j^2$ slices.  Rather, we find that
the relaxation process is able to distribute angular momentum among
the particles, but is unable to produce the predicted distribution.

For simulations 1, 2, 4, 5 \& 6, there is a tidal effect previously
alluded to that may play a role.  Each halo was generated
assuming that it was isolated.  If that were true, the DF would be a {\it
single-valued function} of the total energy alone $f = f(E)$ -- as in
the collapse simulations of section \ref{halo_results1}.  When
placed in proximity to another halo, the net gravitational potential will
no longer be spherically symmetric, and there can, in fact, be a
significant potential gradient across the galaxy.  This has the effect
of changing the total energies of the particles and making $f$ a {\it
multi-valued } as a function of $E$.  Even in the first simulation, when
the galaxies are separated by almost 1 Mpc, the change in potential
across the galaxy is of order $\sim 5\%$.  An energy spread of $5\%$
will give a variation in the DF of $\sim 50\%$ for a weakly bound particle
in the KD model 2 halo we considered (taking $E = -0.1$ in the
expression for $f$).  Even a very tightly bound particle ($E=-10$)
will experience a DF spread of $\sim 10\%$. This leads to a very
complicated initial DF and this may delay the onset of the Gaussian
final state.

However, although the initial multi-valued DF is a major factor
affecting the Gaussian signature, it cannot be the only one and
ultimately it should not prevent it. Thus simulation 3 was
started from initial conditions in which the galaxies were
sufficiently far apart that the spread in the Energy -- DF relation
was small, and yet it also failed to relax to a Gaussian over the
lifetime of the simulation (over 20 Gyr after the galaxies' initial
contact).

Funato, Makino \& Ebisuzaki (1992a, b) also performed a study of
violent relaxation, and found that it proceeds by a combination of
wave--particle interaction and phase mixing.  In the early stage, the
coherent motion of particles causes large-amplitude fluctuations in
the potential field.  The interaction of the individual particles with
this wave can change the energies of the particles significantly in a
classic Landau `damping' mode.   The waves decay rapidly due to the
wave--particle interaction and phase mixing.  In the later stages the
energy change of the particles is much smaller, since the amplitude of
the wave has been significantly reduced.  At this point, relaxation
proceeds primarily through phase mixing (until the phase-mixing
instability becomes important).  Funato et al. show that phase mixing
in this stage of evolution can be slow in the core, and small
oscillations can survive there for a relatively long time (they claim
the oscillations can survive for 10 crossing times or more).

We believe that our simulations also display these oscillations, but
they also show that they are most visible the more finite the number
of particles employed. In the infinite particle limit represented by
our direct integration these fluctuations are smoothed to the point of
invisibility (Fig.\ \ref{plumvirial}) through the addition of very short
wavelengths. Nevertheless the persistence of these oscillations tends
to support the conclusion of Funato et al.  
\begin{figure}
\plotone{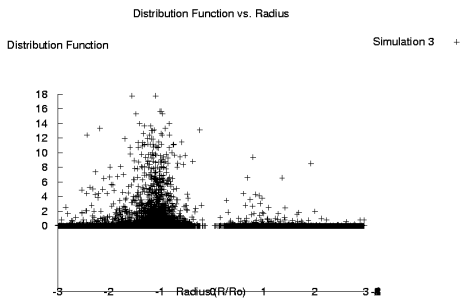}
\caption{\label{halo.collide3.DF.R}
Profile of DF vs.\ radius for colliding galactic haloes (Simulation
3).  This figure clearly shows that the indiviual halo cores have
retained some identity within the merger remnant through the
relaxation process.}
\end{figure}
\nopagebreak
\begin{figure}
\plotone{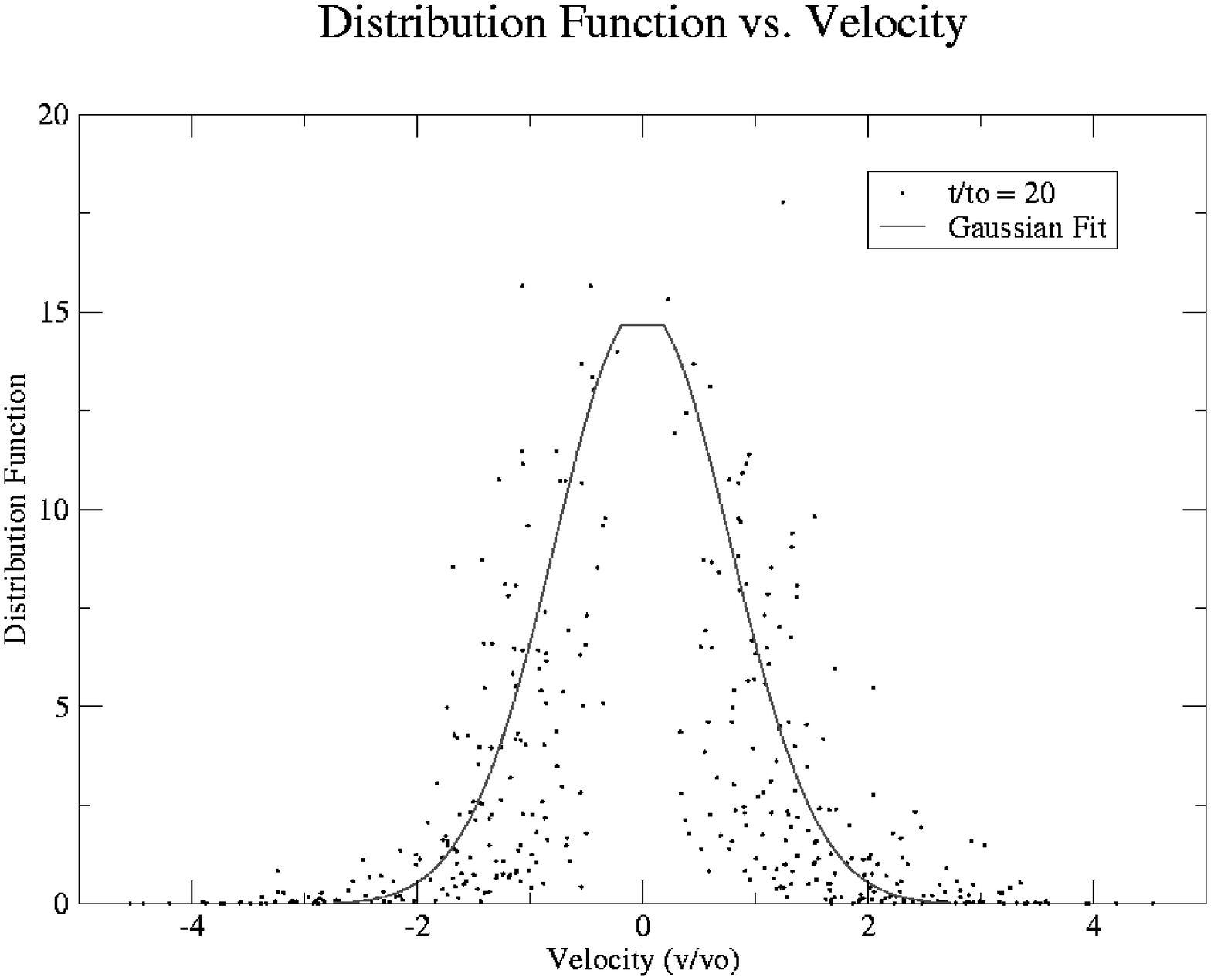}
\caption{\label{halo.collide6.vel}
Profile of DF vs.\ velocity for colliding galactic haloes (Simulation
6).  This slice is taken at a radial bin of width 0.05, centered at
0.025. The velocity distribution appears Gaussian, with the curve
broadened due to perturbation of the DF--Energy relation of the
haloes.}
\end{figure}

In a merger simulation performed by these same authors (Funato, Makino
\& Ebisuzaki 1992b), the cores of the initial galaxies remained
distinct after the galaxies have merged to one remnant. The two cores
oscillate about the center of the remnant. Our $f$($r$) plot for
simulation 3 confirms this observation (Fig.\
(\ref{halo.collide3.DF.R})).  
There are evidently two distinct core
regions in the final remnant as represented by the double-peaked DF.
Funato, Makino \& Ebisuzaki (1992a) note that, ``violent relaxation
disappears within the crossing time scale, no matter whether the
system reaches some equilibrium or not.''  It appears that in the
colliding systems studied here (1,2,3 and 4) the violent relaxation
process does not last long enough to produce the Gaussian velocity
distribution observed in the isolated collapse simulations. Given our
comments above, it is possible that the necessary small scale
interactions have not been sufficiently resolved in these
calculations. Even a reasonable coarse graining would not however
smooth the merger remnant to a convincing Gaussian form.

The initially concentric placement of the haloes in simulation 5 has
the effect of simply doubling the potential energy of each particle
with no increase in its kinetic energy -- essentially generating a
single halo with an initial virial ratio of 0.5, similar to those
studied in the previous section.  Moreover, the effective resolution
is increased. It is therfore not surprising that this simulation shows
the transition to a Gaussian velocity distribution within a few core
orbital periods.

Perhaps our most original result in this section is revealed in
simulation 6. The final velocity distribution of the merger remnant in
simulation 6 {\it is} close to Gaussian in the central region (Fig.\
(\ref{halo.collide6.vel})).  This is true despite the spread in the 
initial DF--Energy relation so that as has already been remarked, this
can not be the dominant factor. Moreover the `relaxation' has occurred
in a relatively short time implying that the violent relaxation has
been relatively efficient while operating over this short time. It
seems then that it more likely to be the case that true collision
mergers require too strong a relaxation mechanism for collisionless
relaxation to be effective as in Funato, Makino \& Ebisuzaki
(1992a). Our new observation is essentially that this lack of
effective relaxation is correlated with the exterior $r^{-3}$ density
profile, while the relaxed systems develop a more nearly NFW type
profile outside the core. Moreover, the relaxed systems have formed by
a process more similar to fragmentation and subsequent re-merging than
to collision of distinct systems. 
\begin{deluxetable}{cll}
\tabletypesize{\scriptsize}
\tablecaption{Collision simulation parameters \label{collision_parameters}}
\tablewidth{0pt}
\tablehead{
\colhead{Label} & \colhead{Initial Separation} & \colhead{Relative Velocity}
}
\startdata
1 & $54 R_o = 0.962$ Mpc & $1.4 v_o = 376.24$\kms \\
2 & $24 R_o = 427.6$ kpc & 0 \\
3 & $140 R_o = 2.5$ Mpc  & $40.3$\kms \\
4 & $4.5 R_o = 80$ kpc   & 0 \\
5 & $0$ kpc              & 0 \\
6 & $1 R_o = 17.8$ kpc   & 0 \\
\enddata
\end{deluxetable}

\section{Discussion and Conclusions}
We have examined in this paper the relaxation of collisionless systems
from non-linearly destabilized initial states toward new equilibrium
states.  Our work combined with previous work indicates that both
violent relaxation (time dependent potential variations and
wave-particle interactions) and unstable phase mixing operate. We have
also confirmed that these processes are not always sufficient to relax
a merging system to a Gaussian DF even at the centre of the system, if
the collision is too energetic. This last statement may be resolution
dependent since there is a hint that with many more particles the
relaxation may be more effective microscopically. However real systems
are finite and will have a `natural' resolution with which they should
be regarded.

We have demonstrated with two entirely independent computational
techniques the approach to a Gaussian velocity distribution of various
collapsing {\it isolated} systems. In one code (CBE integration
method) spherical symmetry is constrained throughout the
collapse,while the N-body tree code does not impose any symmetry.  We
were also able to independently confirm the existence of a radial
phase-mixing instability as described by Henriksen \& Widrow (1997)
{\it using both codes}.  Our results indicate that an isolated halo
will relax toward a Gaussian velocity distribution, with a relaxation
timescale of a few Gyr as predicted by some authors (e.g. Nakamura
(2000)).  This is a fairly robust result since these Gaussian
profiles appear in collapsing polytropes of various indices, and in
destabilized lowered Evans haloes, when two independent numerical
techniques are used. This can happen within a few core orbital periods.

Although several authors (van Albada 1982; Tanekusa 1987;
Funato, Makino \& Ebisuzaki 1992a, b) have disputed this prediction
for various reasons, to our knowledge no one prior to this work has
actually examined the velocity-space structure of the end-state to
search for the Gaussian signature. By plotting the velocity
distributions for different times and radial positions and could
actually observe the evolution of the system to the Gaussian form. We
confirm that violent relaxation can produce the predicted distribution
within a few core orbital periods ((half-mass crossing times in the
case of the polytropic collapse models) -- exactly the time frame over
which violent relaxation is expected to be important. 

Although the systems studied here were collisionless (or approximately
collisionless, in the case of the particle simulations), they were
able to evolve to the predicted Gaussian velocity state.  In the case
of a cold collapse, an instability was observed which it is believed
allows the system to relax by permitting particles to `diffuse' across
characteristics in a less than `finest-graining' representation.  We
expect that the addition of some small collisional interaction will
assist the relaxation (as in Tanekusa 1987) also by allowing particles
to diffuse across characteristics of the CBE. Such a process would
compete with and perhaps suppress the growth of the instability since
the phase streams would no longer be so distinct.

We have identified two effects that lead to requiring longer and
stronger relaxation than is available by collisionless processes to
cause a merger event to forget the initial conditions. One effect is
that the halo models were generated assuming they were gravitationally
isolated (i.e.~no background potential gradient). Thus the DF is a
single-valued function of the total energy $E$.  In the first two
merger simulations, however, the galaxies begin near enough that the
gravitational potential across a galaxy changes by $> \sim 5\%$.  In
this case, the difference in potential across the galaxy is enough to
significantly broaden the relation between $f$ and $E$. 

The more dominant factor which contributes to the incomplete
transition is the limited timescale over which violent relaxation
operates, at least with finite `resolution' (a physical effect in
finite systems). Sufficient initial asymmetry in the merger for a
given number of particles leads to the violent relaxation ending
before the memory independent `relaxed' state is reached. That is the
bulk oscillations implied by the asymmetry do not degrade to `thermal'
motion. We are left  then with a system which has only partially
proceeded to the maximum entropy state. Subsequent mergers might
however restart this process, and if they are gentle enough (or one
system is much smaller than the other) ultimately the system may
relax.  We saw in fact in simulation 6 that systems that are merging
in a closely overlapping state with zero relative velocity do relax to
the Gaussian form with our resolution. This distinction is also
reflected in the final density profiles, wherein the unrelaxed
profiles are approximately  $r^{-3}$ outside the core, while the
relaxed systems have a flat core and what might be described as an
exterior NFW profile.

We conclude then by stating that isolated, finite, collisionless
systems that form by approximately central collapse should relax to a
universal Gaussian form (predicted to be a state of maximum entropy)
when appropriately coarse-grained (that is `resolved' or
`smoothed'). However, those that form through the merging of systems
of comparable size will not relax, given the same coarse-graining. 

Our present simulations do not start from strict cosmological
conditions (although the cold collapses are not so very different) and
in particular the initial density profiles have flat cores. Thus we
can not comment directly on whether or not the Cold Dark Matter (CDM)
heirarchical halo formation theory leads to central cusps or
cores. The Gaussian velocity distribution is also compatible with a
singular isothermal sphere and the corresponding $r^{-2}$
profile. However one might expect that this isolated solution is not
the maximum entropy limit and that in fact the Gaussian distribution
corresponds in general to a flat core. Haloes that have not relaxed to
the Gaussian state may well show cusp profiles, especially if they
have begun in this fashion (unlike our examples). Our results allow us
to speculate however that the accretion of a dwarf halo by an
unrelaxed massive halo may restart the relaxation process in the large
halo. This might lead through a series of such episodic relaxation
events to the formation of a flat core. We intend to investigate this
speculation in future work.

\acknowledgments
TECM wishes to acknowledge the financial support of Queen's University
and the Ontario Graduate Scholarship in Science and Technology
program. RNH is grateful for a research grant from the Canadian
Natural Sciences and Engineering Research Council.


\end{document}